\documentclass[manuscript,screen]{acmart}

\AtBeginDocument{%
  \providecommand\BibTeX{{%
    \normalfont B\kern-0.5em{\scshape i\kern-0.25em b}\kern-0.8em\TeX}}}

\acmJournal{JACM}
\acmVolume{37}
\acmNumber{4}
\acmArticle{111}
\acmMonth{8}

\acmISBN{978-1-4503-XXXX-X/18/06}

\usepackage{makecell}
\usepackage{algorithm}
\usepackage{algorithmicx}
\usepackage{algpseudocode}
\usepackage{spverbatim}
\usepackage{fancyvrb}
\usepackage{fvextra}
\usepackage{subcaption}
\usepackage{halloweenmath}

\usepackage[utf8]{inputenc}
\DeclareUnicodeCharacter{2212}{-}

\begin{document}

\title{HW-V2W-Map: Hardware Vulnerability to Weakness Mapping Framework for Root Cause Analysis with GPT-assisted Mitigation Suggestion}


\authornote{Authors contributed equally to this research.}

\author{Yu-Zheng Lin}
\authornotemark[1]
\email{yuzhenglin@arizona.edu}
\affiliation{%
  \institution{University of Arizona}
  \streetaddress{1230 E Speedway Blvd}
  \city{Tucson}
  \state{Arizona}
  \country{USA}
  \postcode{85721}
}

\author{Muntasir Mamun}
\authornotemark[1]
\email{muntasir@arizona.edu}
\affiliation{%
  \institution{University of Arizona}
  \streetaddress{1230 E Speedway Blvd}
  \city{Tucson}
  \state{Arizona}
  \country{USA}
  \postcode{85721}
}

\author{Muhtasim Alam Chowdhury}
\authornotemark[1]
\email{mmc7@arizona.edu}
\affiliation{%
  \institution{University of Arizona}
  \streetaddress{1230 E Speedway Blvd}
  \city{Tucson}
  \state{Arizona}
  \country{USA}
  \postcode{85721}
}

\author{Shuyu Cai}
\authornotemark[1]
\email{sycai@ucdavis.edu}
\affiliation{%
  \institution{University of California Davis}
  \streetaddress{1 Shields Ave}
  \city{Davis}
  \state{California}
  \country{USA}
  \postcode{95616}
}

\author{Mingyu Zhu}
\authornotemark[1]
\email{myuzhu@ucdavis.edu}
\affiliation{%
  \institution{University of California Davis}
  \streetaddress{1 Shields Ave}
  \city{Davis}
  \state{California}
  \country{USA}
  \postcode{95616}
}

\author{Banafsheh Saber Latibari}
\authornotemark[1]
\email{bsaberlatibari@ucdavis.edu}
\affiliation{%
  \institution{University of California Davis}
  \streetaddress{1 Shields Ave}
  \city{Davis}
  \state{California}
  \country{USA}
  \postcode{95616}
}

\author{Kevin Immanuel Gubbi}
\authornotemark[1]
\email{kgubbi@ucdavis.edu}
\affiliation{%
  \institution{University of California Davis}
  \streetaddress{1 Shields Ave}
  \city{Davis}
  \state{California}
  \country{USA}
  \postcode{95616}
}

\author{Najmeh Nazari Bavarsad}
\authornotemark[1]
\email{nnazari@ucdavis.edu}
\affiliation{%
  \institution{University of California Davis}
  \streetaddress{1 Shields Ave}
  \city{Davis}
  \state{California}
  \country{USA}
  \postcode{95616}
}

\author{Arjun Caputo}
\email{acaputo@arizona.edu}
\affiliation{%
  \institution{University of Arizona}
  \streetaddress{1230 E Speedway Blvd}
  \city{Tucson}
  \state{Arizona}
  \country{USA}
  \postcode{85721}
}
\author{Avesta Sasan}
\email{asasan@ucdavis.edu}
\affiliation{%
  \institution{University of California Davis}
  \streetaddress{1 Shields Ave}
  \city{Davis}
  \state{California}
  \country{USA}
  \postcode{95616}
}

\author{Houman Homayoun}
\email{hhomayoun@ucdavis.edu}
\affiliation{%
  \institution{University of California Davis}
  \streetaddress{1 Shields Ave}
  \city{Davis}
  \state{California}
  \country{USA}
  \postcode{95616}
}

\author{Setareh Rafatirad}
\email{srafatirad@ucdavis.edu}
\affiliation{%
  \institution{University of California Davis}
  \streetaddress{1 Shields Ave}
  \city{Davis}
  \state{California}
  \country{USA}
  \postcode{95616}
}

\author{Pratik Satam}
\email{pratiksatam@arizona.edu}
\affiliation{%
  \institution{University of Arizona}
  \streetaddress{1230 E Speedway Blvd}
  \city{Tucson}
  \state{Arizona}
  \country{USA}
  \postcode{85721}
}

\author{Soheil Salehi}
\email{ssalehi@arizona.edu}
\affiliation{%
  \institution{University of Arizona}
  \streetaddress{1230 E Speedway Blvd}
  \city{Tucson}
  \state{Arizona}
  \country{USA}
  \postcode{85721}
}

\renewcommand{\shortauthors}{Yu-Zheng Lin, Muntasir Mamun, Muhtasim Alam Chowdhury, et al.}

\begin{abstract}
\textbf{Abstract:} 
The escalating complexity of modern computing frameworks has resulted in a surge in the cybersecurity vulnerabilities reported to the National Vulnerability Database (NVD) by practitioners. Despite the fact that the stature of NVD is one of the most significant databases for the latest insights into vulnerabilities, extracting meaningful trends from such a large amount of unstructured data is still challenging without the application of suitable technological methodologies. Previous efforts have mostly concentrated on software vulnerabilities; however, a holistic strategy incorporates approaches for mitigating vulnerabilities, score prediction, and a knowledge-generating system that may extract relevant insights from the Common Weakness Enumeration (CWE) and Common Vulnerability Exchange (CVE) databases is notably absent. As the number of hardware attacks on Internet of Things (IoT) devices continues to rapidly increase, we present the Hardware Vulnerability to Weakness Mapping (HW-V2W-Map) Framework, which is a Machine Learning (ML) framework focusing on hardware vulnerabilities and IoT security. The architecture that we have proposed incorporates an Ontology-driven Storytelling framework, which automates the process of updating the ontology in order to recognize patterns and evolution of vulnerabilities over time and provides approaches for mitigating the vulnerabilities. The repercussions of vulnerabilities can be mitigated as a result of this, and conversely, future exposures can be predicted and prevented. Furthermore, our proposed framework utilized Generative Pre-trained Transformer (GPT) Large Language Models (LLMs) to provide mitigation suggestions.

\end{abstract}




\authorsaddresses{} 
\maketitle

\section{Introduction}
The growing complexity of modern computer systems is increasingly susceptible to ever-evolving, highly sophisticated hardware and software security attack vectors. One such evolution of attack vectors is software-assisted hardware attacks which exploit the computing system's hardware weaknesses to expose vulnerabilities that can have catastrophic impacts \cite{kim2014flipping, yarom2014flush+, kocher2020spectre, liu2015last, gruss2016flush+, he2017secure}. An example of a software-assisted hardware attack is the \texttt{Row-Hammer (RH)} attack \cite{kim2014flipping} wherein the attacker's frequent accessing of select DRAM rows causes bit flips in nearby rows which then opens up vulnerabilities such as privilege escalation and data stealing. Moreover, the increasing density of memory chips, due to the packing of memory cells, intensifies the impact of such attacks resulting in the decrease of the \texttt{Hammer Threshold ($T_{RH}$)}  from 139K in DDR3 DRAM in 2014 to  4.8K LPDDR4 in 2020.  This general trend of escalation in attack impact and ease of execution can be seen for all software-assisted hardware attacks, highlighted by the increase in the number of Common Vulnerabilities and Exposure (CVE) entries in the National Vulnerability Database (NVD) with 17,302 entries reported in 2019\cite{NVD} \cite{MITRE}. The McAfee threats report for the first quarter of 2020 recorded 375 threats every minute \cite{Labs_2020}; a 71\% increase in new mobile malware reporting during the first quarter of 2020, showing a 12\% escalation compared to the preceding four quarters. The ever-increasing demand for mobile/edge computing and IoT devices, the scarcity of onboard computational resources, and the need to keep development costs low, increase the challenges in securing such devices. It is anticipated that this trend will worsen with the addition of 75 Billion Internet of Things (IoT) devices by the year 2025  \cite{zhou2018effect}, allowing attackers to execute even more sophisticated, coordinated, and stealthy cyber attacks \cite{zhou2018effect, yang2017hardware, tsang2022fandemic, kolhe2021securing, kolhe2022lock, hassan2023circuit, shukla2019rnn, shukla2019stealthy, salehi2022neuromorphic}. \\
In response to these issues, we employ Machine Learning (ML) and Natural Language Processing (NLP) methods to analyze the Internet of Things (IoT) hardware vulnerabilities and weaknesses found in the Common Vulnerabilities and Exposures (CWEs) and Common Vulnerabilities and Exposures (CVEs) databases. We also look into the relationships between these vulnerabilities and weaknesses, identify trends, and provide vulnerability mitigation strategies, using Generative Pre-trained Transformer (GPT) Large Language Models (LLM). The following are the primary contributions of this paper:
\begin{enumerate}
 \item We take a conceptual modeling approach to the development of our hardware-security ontology. This approach helps analyze software-assisted hardware attack types, patterns, and their evolution.
 \item Using user queries, our interactive framework discovers patterns and linkages between different vulnerabilities and their impacts on IoT. This analysis uses  NLP techniques like n-grams and part-of-speech (POS) tagging, to process the available data to extract and present the user with a subset of pertinent and useful articles related to their query.
 \item An interactive Graphical User Interface (GUI) is proposed for our suggested framework, which is capable of identifying vulnerability mitigation strategies, storytelling, and the development of information or knowledge generation.
 \item Exploit and impact scores are determined using a machine learning approach that relies on inputs from the GUI, accomplished by constructing a Common Vulnerability Scoring System (CVSS) threat vector containing base score metrics and data cleaning. 
 \item We developed a feature called ``GPT-assisted Mitigation Suggestion.'' After deriving the enumerated relevant CWE vulnerabilities, the system will use a web crawler to obtain potential mitigation information from the CWE website and integrate the vulnerability description entered by the user. This information will be embedded in our prompt engineering. Users will get mitigation measures provided by OpenAI's GPT model.
\end{enumerate}

\begin{table}[!t]
	\caption{A comparison of our suggested framework with other similar tools that are considered to be state-of-the-art is presented.}
	\centering
	\resizebox{\columnwidth}{!}{
		\begin{tabular}{ccccccccccc}
			\hline
			\hline
                    Methodology & \thead{Data Source}    & Scope   & IoT        & \thead{Storytelling} & Automated & \thead{Machine \\ Learning}    & \thead{Knowledge \\ Generation}  & \thead{Mitigation \\ Suggestion} & \thead{Large Language\\ Model (GPT)} \\
			\hline
			\hline
			\cite{li2019comparative} Li et al.         & NVD+SARD      & Software & X          & X            & \checkmark & \checkmark & X          & X    & X      \\
			\hline
			\cite{neuhaus2010security} Neuhaus et al.    & NVD           & Software & X          & X            & \checkmark & X          & X          & X     & X     \\
			\hline
			\cite{blinowski2020cve} Blinowski et al.  & NVD           & Software & \checkmark & X            & X          & X          & X          & X      & X    \\
			\hline
			\cite{murtaza2016mining} Murtaza et al.    & NVD           & Software & X          & X            & X          & X          & X          & X     & X     \\
			\hline
			\cite{williams2018analyzing}Williams et al.    & NVD           & Software & X          & \checkmark   & \checkmark & X          & X          & X    & X      \\
			\hline
			\cite{hassan2023automated}Hassan et el.      & NVD+CWE       & Hardware & \checkmark & \checkmark   & \checkmark & \checkmark & X          & X     & X     \\
			\hline
			\cite{kuehn2021ovana}Kuhen et el.       & NVD           & Software & X          & X            & \checkmark & \checkmark & X          & X     & X     \\
			\hline
			\cite{janiszewski2021automatic}Janiszewski et al. & BID+CNNVD+NVD & Software & \checkmark & X            & X          & X          & \checkmark & X     & X     \\
			\hline
			\cite{guo2021key} Guo et al.        & IBM X-Force   & Software & X          & X            & \checkmark & \checkmark & \checkmark & X      & X    \\
			\hline
			Proposed HW-V2W-Map        & NVD+CWE       & Hardware & \checkmark & \checkmark   & \checkmark & \checkmark & \checkmark & \checkmark & \checkmark \\
			\hline     
			\hline
		\end{tabular}
	}
	\label{tab:example}
\end{table}

The remaining paragraphs of this document are structured as described below.
A brief background study is presented in Section 2. An explanation of the databases that were utilized in the course of our research and analysis is provided in Section 3. In Section 4, we detail our suggested methodology to employ NLP techniques to extract information that is both valuable and understandable from the databases that were discussed in Section 3. This section presents the outcomes of the approach that we have proposed and explores the significance of the findings that we have obtained. Section 5 discusses the findings and why they matter. Section 6 presents practical ways to incorporate GPT LLMs to provide mitigation for the vulnerabilities. Lastly, Section 7 sums up the advantages of their approach and highlights the main points. 

\section{Background Study}
While recent computing systems are optimized for improved performance, there has been an escalation in the prevalence of security attacks, attributable to an enlarged attack surface introduced by an increase in heterogeneity and complexity \cite{nistSoftwareAssurance}. This work aims to provide a framework to more effectively recognize software-assisted hardware vulnerabilities susceptible to exploitation and to provide defensive strategies, as the behavior of such vulnerabilities has only recently been attempted to be understood in recent studies.  Developing effective solutions to safeguard hardware devices from existent CWE classes of vulnerabilities that are exploitable is critically essential and significant. Two different datasets were collected from different applications by Li et al. in 2019 \cite{li2019comparative}, which included 126 different types of vulnerabilities. The initial comparative analysis, which attempted to statistically examine the effects of multiple elements on the effectiveness of vulnerability identification was carried out with the use of these datasets. A classification system of vulnerability data related to the Internet of Things (IoT) and Industrial Internet of Things (IIoT) devices is proposed by Blinowski et al. \cite{blinowski2020cve}. A methodology that is based on machine learning was presented by Hassan et al.\cite{hassan2023automated} to classify vulnerabilities and their impact vectors. The author focused on hardware vulnerabilities in the Internet of Things domain within this methodology. The Ontology-driven Storytelling Framework (OSF) within the framework seeks to detect recurring patterns in vulnerabilities. This is accomplished with the assistance of an automatic ontology updating process. Blinowski et al. \cite{blinowski2020cve} classified the CVE records from the NVD database into seven distinct categories based on the breadth of the Internet of Things applications. Additionally, the writers tagged the database samples based on their expert knowledge.
Once the labeling process was completed, the authors utilized a Support Vector Machine (SVM) classifier on the device and vulnerability data.  This allows the authors to anticipate, and when feasible, avert and mitigate attacks that are caused by new vulnerabilities.\\

Williams et al. \cite{williams2018analyzing} presented an integrated data mining system that can automatically illustrate the progression of vulnerabilities over time and identify the development of a particular cybersecurity attack. In \cite{williams2018analyzing}, a Supervised Topical Evolution Model (STEM) was employed by the authors, a model specifically developed for deriving time-related patterns from a text corpus. This model's purpose is to uncover patterns in vulnerabilities by tapping into the latent framework of the text corpus and merging the time-related elements of vulnerability reports. Following that, they employed a storytelling approach that is based on diffusion, which involves searching through previous vulnerability reports in order to depict how a present threat has developed over time. Additionally, they examined the progression of vulnerabilities, tracing their development through vulnerability records. In addition, Neuhaus et al. \cite{neuhaus2010security} offered a classification strategy that makes use of Latent Dirichlet Allocation (LDA) on the descriptive texts of CVE entries. The purpose of this approach was to construct a classification system and automatically detect common subjects and developing patterns. Within the context of software applications, Murtaza et al. \cite{murtaza2016mining} studied the utilization of historical patterns of vulnerabilities as a means of forecasting future vulnerabilities, wherein the authors assert that they have not discovered any statistically significant information regarding the patterns of vulnerability occurrence over time. An intriguing discovery made by this research is that the progression of vulnerability events follows a first-order Markov process, allowing forecasting the subsequent vulnerability by relying solely on the prior vulnerability. To evaluate the quality of vulnerability databases, Kuhen et al. \cite{kuehn2021ovana} developed Overt Vulnerability source ANAlysis (OVANA), a system that uses a combination of NLP and ML to search unstructured text for details absent from structured fields, to improve the vulnerability database with the newly discovered information. Research has shown that OVANA improves the NVD by 51.23 percent. Similarly, Janiszewski et al. \cite{janiszewski2021automatic} present a Vulnerability and Attack Repository for the Internet of Things (VARIoT), a standardized repository of IoT vulnerabilities wherein AI-based trust management and meta-information extraction is implemented, aiming to improve the completeness, usability, and dependability of the data. An investigation was carried out by Guo et al. \cite{guo2021key} to determine whether or not it is capable of effectively predicting the severity of a security vulnerability by completing the components of the vulnerability that are pertinent to the vulnerability. Xiaozhou et al.\cite{li2023anatomy} worked on a systematic review study and detailed analysis, where they found (1) the prevalent vulnerability databases that have been utilized; (2) the objectives behind their adoption; (3) the additional information sources that have been adopted; (4) the methodologies and approaches employed; and (5) the tools suggested for understanding insight into the structure of existing vulnerability databases. A deeper comprehension of these factors may help practitioners create trustworthy information sources to guide their security standards and policies, as well as researchers make well-informed selections about which databases to use while conducting research.

Coupled with the rising significance of large language models (LLMs) in cybersecurity, their findings provide a compelling basis for exploring the potential synergy between LLMs and the proposed framework in addressing hardware vulnerabilities. In terms of cyber security research with a language model, Ziems et al. \cite{ziems2021security} addressed the task of identifying software vulnerabilities by treating it as a natural language processing (NLP) problem, where source code is treated as textual data. They curated a dataset comprising over 100,000 C programming code files covering 123 distinct vulnerability categories. The researchers pre-processed the NIST NVD/SARD databases for training and testing purposes. Their approach involved utilizing BERT with Bidirectional LSTM, demonstrating its effectiveness in the detection of software vulnerabilities. In recent years, within the domain of natural language processing, researchers have tried to increase the size of training vocabulary and models to solve a wide range of tasks\cite{wei2022emergent}. This methodology has been coined as the ``Large Language Model'' (LLM) approach. Large language models are trained on diverse datasets. They can perform various NLP tasks, not only complete writing, translation, and programming, but also have potential applications in advanced scientific fields, such as medicine, materials science\cite{jablonka202314}, and genetics\cite{duong2023analysis}. Currently, Big Tech companies such as Microsoft, Meta, and Google have released or invested in several well-known LLM model services, such as GPT-4 \cite{openaiGPT4}, LLaMA\cite{metaIntroducingLLaMA}, Bard \cite{blogIntroducingPaLM}, etc. The application of LLMs presents emerging research opportunities in the scope of cybersecurity research. Pearce et al.\cite{10179324} investigated the utilization of LLMs for zero-shot vulnerability repair with Codex and Jurassic J-1. By modifying the prompts, LLMs can provide feedback on the repaired version of the insecure code. Al-Hawawreh et al.\cite{al2023chatgpt} explored potential applications of ChatGPT in the field of cyber security, encompassing various domains such as Honeypots, Code Security, Malware Development, Phishing and Social Engineering, Cybersecurity Reports and Consulting, Vulnerability Scanning and Exploitation, Disinformation and Misinformation, and Cybersecurity Education. Also, they tried to conduct false data injection attacks on industrial systems. The experiments showed that ChatGPT provided a simulation code for closed-loop control and successfully injected inaccurate data into the closed-loop control system. The ChatGPT also created an anomaly detection code based on isolation Forest. These studies illustrate the potential utility of LLMs in cybersecurity research.

The majority of the previous works have decided to concentrate their research on software vulnerabilities. Moreover, most of the previous works, despite the significant amount of effort that they have put in, have not been able to provide a framework for mitigation techniques that are capable of uncovering insights about the relationship and trends that have occurred within the CVE and CWE databases over time. Furthermore, these works have not been able to provide detailed insights into how vulnerability and impacts have developed over time. With a particular emphasis on vulnerabilities that are associated with Internet of Things (IoT) hardware, the purpose of this article is to investigate the patterns and connections that have emerged over a period of time. The approach that we have developed aims to discover comparable patterns of vulnerabilities throughout time. This can assist in mitigating the effects of vulnerabilities or in predicting and preventing problems in the future. This type of study of the similarities between vulnerabilities provides a vital understanding of whether the patterns rising or declining. The inclusion of LLMs with the proposed framework could act as an assistant to provide potential mitigation information related to hardware vulnerabilities. The innovative nature of our proposed framework is highlighted in Table I, which compares it to the current state of the art.

\section{Security Vulnerability Information:}

\subsection{National Vulnerability Database (NVD):} 
The National Vulnerability Database (NVD) \cite{NVD} is a repository for vulnerability management data of the United States government, built using the Security Content Automation Protocol (SCAP), aiming to automate vulnerability management, security measurement, and compliance. The NVD content is continuously updated through data feeds in JSON format and includes data points that are added when new information about a particular threat becomes available. Our framework parses the NVD to extract CVE entries, using this data for vulnerability analysis and ontology-based mitigating techniques, answering the queries prompted by user input. 

\subsubsection{Principal Elements and Characteristics of NVD}
\begin{itemize}
\item \textbf{Comprehensive Vulnerability Database:} NVD is responsible for the upkeep of a comprehensive database that is constantly updated. This database covers vulnerabilities that are present in a wide range of technologies, including software, hardware, and firmware.
\item \textbf{Integration of CVE:} NVD makes use of the Common Vulnerabilities and Exposures (CVE) system, which guarantees a naming convention that is standardized and recognized all around the world for vulnerabilities that have been discovered. Each vulnerability is given a CVE identifier that is unique to it.
\item \textbf{Structured Vulnerability Data:} The information on vulnerabilities is supplied in a structured fashion, offering details such as the products that are affected, severity scores (CVSS), possible impact, and references to advisories and solutions that are connected to the vulnerabilities.
\item \textbf{Searchable and Accessible Platform:} The NVD provides a searchable interface that security professionals, researchers, and the general public can utilize to reach vulnerability information efficiently. In order to facilitate individualized searches, the platform allows users to conduct queries based on a variety of parameters.
\item \textbf{Common Vulnerability Scoring System (CVSS):} The Common Vulnerability Scoring System (CVSS) is utilized to evaluate the severity of vulnerabilities, offering a standardized framework for assessing the effect of vulnerabilities as well as their potential to be exploited. CVSS score consists of Base, Temporal, and Environmental metrics, where the Base metrics measure the intrinsic qualities of the vulnerability, the Temporal metrics evaluate measures that change over the vulnerability's lifetime, especially as it gets disclosed and mitigation techniques are designed, and the Environmental metrics captures the environmental factors affecting the vulnerability.
\item \textbf{Data Feeds and APIs:}  The NVD offers data feeds and APIs that organizations can use to integrate vulnerability information into their security tools and processes. Both the automation and the timely disclosure of vulnerability data are made easier as a result of this.
\end{itemize}

\subsubsection{Role of NVD in Vulnerability Assessment:}
\begin{itemize}
\item \textbf{Early Warning and Threat Intelligence:}  NVD functions as an early warning system, delivering information promptly on newly found vulnerabilities. Consequently, this makes proactive threat intelligence more accessible and assists organizations in remaining one step ahead of potential hazards.
\item \textbf{Risk Management:} To properly identify and manage cybersecurity risks, organizations use data from NVD. They can apply mitigation techniques and prioritize security activities if they are thoroughly aware of the vulnerabilities linked to their systems.
\item \textbf{Coordination and Collaboration:}  The non-governmental organization (NVD) encourages collaboration between various players in the sector, government agencies, and the worldwide cybersecurity community. It encourages information sharing and collaborative efforts to address security concerns by centralizing vulnerability information, which in turn promotes information sharing.
\item \textbf{Research and Analysis:} Cybersecurity experts and analysts use the NVD to conduct in-depth studies and analyze vulnerabilities. Developing insights into new risks, attack patterns, and potential solutions is made possible with the assistance of the database.
\item \textbf{Recommendations and Best Practises:} NVD entries frequently include recommendations for best practices, mitigation, and instructions for addressing identified vulnerabilities. The development of efficient security plans is facilitated by the inclusion of this information in organizations.
\end{itemize}

However, continuous attempts are being made to improve the NVD in order to better its performance in terms of functionality, accuracy, and relevance. Continual research, feedback from stakeholders, and collaboration with the cybersecurity community are all factors that contribute to the enhancement of the database's capabilities.

\subsection{Common Weakness Enumeration (CWE) Database:} 
The Common Weakness Enumeration Database, often known as the CWE Database\cite{MITRE}, systematically classifies and enumerates common vulnerabilities that are present in both software and hardware installations, standardizing the discussion, categorization, and management of all vulnerabilities. This framework plays an essential part in improving software application security posture, connecting the fundamental causes of each vulnerability's occurrence. In this work, we use the CWE, where the proposed framework uses data from CSV files in the hardware navigation view, which is available for public distribution. We use this community-contributed database to characterize the weaknesses in the linkages that our ontology has created. We leverage the hardware design perspective, which contains more than 100 hardware weaknesses linked to each hardware-based IoT CVE with its matching hardware weakness CWE. 

\subsubsection{Principal Elements and Characteristics of CWE Database}
\begin{itemize}
\item \textbf{Comprehensive Weakness Catalogue:} CWE keeps an extensive list of typical software security flaws that span a variety of frameworks, programming languages, and development techniques.
\item \textbf{Common Identifiers:}  Every vulnerability that is stored in the CWE Database is given a one-of-a-kind identification, which guarantees that the cybersecurity community will adhere to a naming practice that is consistent throughout. These identifiers are required to promote clear communication and an understanding of shortcomings.
\item \textbf{Structured Taxonomy:}  To categorize weaknesses, the database makes use of a taxonomy arrangement that is both hierarchical and structured. The organization and navigation of weaknesses based on their characteristics and relationships are made easier as a result of this.
\item \textbf{Mitigation and Best Practice:}  CWE provides comprehensive information about each vulnerability, including examples of potential mitigation, best practices, and recommended methodologies for developing secure software. Because of this, developers and security professionals are given the ability to address vulnerabilities proactively.
\item \textbf{Mapping to Other Standards:}  Common Vulnerabilities and Exposures (CVE) and Common Attack Pattern Enumeration and Classification (CAPEC) are two examples of industry standards that are aligned with and complement the Common Web Encryption standard. The effectiveness of the interoperability across various cybersecurity frameworks is improved by this integration \cite{SAUERWEIN2019140}.
\item \textbf{Community Contributions:} The CWE Database actively encourages contributions from members of the cybersecurity community, such as researchers, practitioners, and industry professionals. This collaborative method guarantees that the database will continue to be dynamic and accurately reflect the ever-changing landscape of software security.
\end{itemize}

\subsubsection{CWE Database for Root Cause Analysis}
\begin{itemize}
\item \textbf{Guidance on Secure Coding:} The CWE is an invaluable resource for developers who are looking for guidance on secure coding practices. In order to proactively address potential security problems, developers might refer to weaknesses that have been found and the recommended practices that are linked with them.
\item \textbf{Risk Assessment and Management:} Companies use CWE to manage the security of their software portfolios and to conduct risk assessments within their organizations. When it comes to effective risk mitigation and safe software deployment, having an understanding of typical weaknesses and fixing them is a contributing factor.
\item \textbf{Education and Training:}  CWE is an essential component of education and training programs centered on developing secure software. Through the provision of a standardized vocabulary for the discussion of security issues, it helps to promote a common understanding among development teams.
\item \textbf{Integration of Tools:} Static analysis scanners and security tools integrate CWE identifiers in order to notify and prioritize vulnerabilities that have been discovered. During the development life cycle, this integration makes the process of identifying and fixing vulnerabilities more transparent and efficient.
\item \textbf{Compliance and Assurance:}  CWE supplies a reference for secure coding standards, which helps with compliance efforts and ensures such efforts are successful. CWE can serve as a benchmark for organizations to evaluate how well they adhere to the prevalent standards and best practices in their industry.
\end{itemize}

Improvements are made to the CWE Database in order to integrate feedback from the cybersecurity community, address newly discovered vulnerabilities, and update existing entries. Using this iterative method guarantees that CWE will continue to be a dependable and relevant resource for the development of safe software.

\section{Methodology}
This section aims to discuss the principles of structuring our data, as well as the algorithms and models we have developed.

\subsection{Proposed Modeling method:} 
This work analyzes the data from the CWE and the CVE using NLP techniques like n-grams, and part-of-speech (POS) tagging\cite{rahman2023attackers}, to construct a descriptive model, that is used for ontology-based storytelling; where the ontology serves users as a model for runtime information exploration where knowledge is represented in the form of subject-predicate-object triples \cite{pan2022ontology}. The constructed ontology facilitates exploring CWE and CVE entities by navigating through the predicates (i.e., relationships) linking the entities in the corresponding knowledge graph\cite{sikos2023cybersecurity}. To construct an ontology in this work, we extracted bigrams, trigrams, and quadgrams from the CVE and CWE databases by utilizing the built-in n-gram function from the NLTK package \cite{bird2009natural}. Our ontology construction algorithm is illustrated in \ref{alg:topic-modeling}, describing the workflow of our proposed topic modeling approach.
During our experiments with the CVE data, we observed that the target of the vulnerabilities is either corporations or names that may typically be identified as people. Therefore, we used PoS tagging from the SpaCy NLP package \cite{honnibal2017spacy} to perform Named-Entity-Recognition as a technique to identify the target of an exploit. This process involves locating all instances of `PERSON' and `ORG' in the beginning. A fail-safe measure that we discovered after evaluating all of the vulnerability descriptions was to obtain the NN noun - singular and the NNS noun - plural from the NLTK POS-tagger. This approach aims to identify the noun, most likely representing the target of an exploit. Identifying the impact of each vulnerability is an additional input required during the process of building the ontology. Upon applying the Collocations and Association Measures function, we ranked the scores and retrieved the associations and word occurrence relationships to better describe the vulnerabilities' impacts.  Our further investigations revealed that trigrams and quadgrams offered a more accurate semantic representation of the vulnerability.

\begin{algorithm}[t!] 
\caption{Modeling the entities of the suggested method}\label{alg:topic-modeling}
\footnotesize
\begin{algorithmic}[1]
\State \textbf{Input}: $nvd\_cwe\_data$
\State \textbf{Output}:$list\_relevant\_objects$
\Function{ngrams}{nvd\_cwe\_data}
    \State tokens := tokenize (nvd\_cwe\_data)
    \State bigr, trgr, quadgr := form\_ngrams (tokens)
    \State \textbf{return} big, trgr, quadgr
\EndFunction
\Function{exploit\_target}{bigr, trgr, quadgr}
\If{cpe\_disctionary$\exists$ bigr, trgr, quadgr}
    \State \textbf{return} cpe\_match
    \ElsIf{entity\_label\_ORG $\exists$ bigr, trgr, quadgr}
    \State \textbf{return} ORG
    \ElsIf{entity\_label\_PERSON $\exists$ bigr, trgr, quadgr}
    \State \textbf{return} PERSON
    \Else
    \State bgs, tgs, qgs := max\_freq\_dist (most\_common(bigr, trgr, quadgr))
    \For{\texttt{<x in bgs, tgs, qgs>}}
        \State tagged=pos\_tag(x)
        \If{NN $\lor$ NNS $\in$ tagged}
            {return} NN $\lor$ NNS
            \EndIf
        \EndFor
    \EndIf
\EndFunction

\Function{attack\_impact}{bigr, trgr, quadgr}
    \State bg\_score,tg\_score,qg\_score := collect\_assc\_meas (bigr, trgr, quadgr)
    \State pmi = nbest\_likelihood\_ratio (bg\_score, tg\_score, qg\_score)
    \State \textbf{return} $max(pmi)$
\EndFunction
\Function{string\_clean}{input\_string}
    \State input\_string := remove\_stopwords,remove\_punctuation, remove\_nonsensical, tokenize,stemming (input\_string)
    \State \textbf{return} input\_string
\EndFunction
\Function{cosine\_similarity}{clean\_string, exploit\_target, attack\_impact}
    \For{\texttt{<x in exploit\_target, attack\_impact>}}
        \State similarity\_score, index\_cve := cosine\_distance (clean\_string, x)
        \State frame\_cve $+=$ (similarity\_score, index\_cve)\EndFor
    \State \textbf{return} max (frame\_cve)
\EndFunction
\State bigr, trgr, quadgr := ngrams (nvd\_cwe\_data)
\State targets := exploit\_target (bigr, trgr, quadgr)
\State impacts := attack\_impact (bigr, trgr, quadgr)
\State clean\_string := string\_clean (input\_string)
\State list\_relevant\_objects := cosine\_similarity (clean\_string, targets, impacts)
\State \textbf{return} (list\_relevant\_objects)
\end{algorithmic}
\end{algorithm}

\subsection{Automated Topic Modeling Framework and Proposed Hardware-Security Ontology} 
\begin{figure}[t!]
\centering
\includegraphics[width=\linewidth]{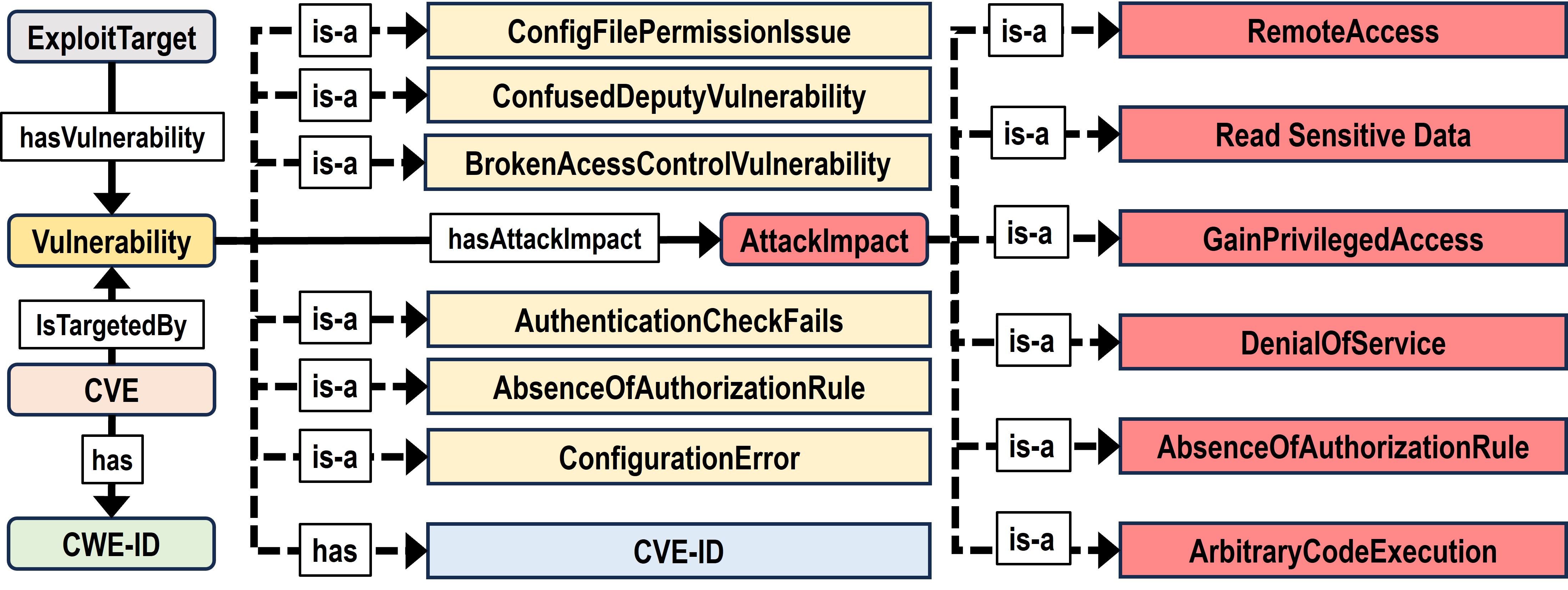}
\caption{Hardware Security Ontology Framework.}
\label{fig:relationship}
\end{figure}

The proposed framework builds an ontology of hardware vulnerabilities, a comprehensive knowledge base in the field of cyber-physical systems (CPS) security. Our main focus is on how to methodically investigate a set of vulnerabilities and construct a story that illustrates the possible effects a vulnerability could have on a CPS system. To facilitate the development of the ontology, we introduce four distinct concepts: Vulnerability, CWE, AttackImpact, and ExploitTarget. Here, we offer concise descriptions of these ontology classes:
\begin{enumerate}
\item Vulnerability: In the complex world of hardware design, errors can occur due to several flaws such as design flaws, manufacturing flaws, or intentional faults (in terms of hardware Trojan). A vulnerability refers to a flaw in the hardware system that could be exploited by malicious actors.
\item AttackImpact: When a vulnerability is exploited by an attacker, it inevitably leads to an impact on the system. The AttackImpact class encapsulates the concept of potential system impacts resulting from such attacks.
\item CWE: The CWE class denotes the specific weakness type associated with a particular vulnerability. 
\item ExploitTarget: A victim system may possess vulnerabilities that could potentially be targeted by attackers for exploitation, resulting in harm to the victim. These specific targets are defined as ExploitTargets.
\end{enumerate}

We examine some important object properties that represent the relationships between the above classes in our ontology:
\begin{itemize}
\item  Exploits: The Exploits object property connects each vulnerability that aims to exploit an ExploitTarget class.
\item hasAttackImpact: Linking the AttackImpact and ExploitTarget classes, this attribute indicates that a system with one or more ExploitTarget objects could be subject to different consequences if a vulnerability is exploited.
\item TargetsCWE: Each vulnerability is associated with a weakness type identified by a CWE-ID, which is represented through the TargetsCWE relationship.
\end{itemize}

We use predicate logic to formally express the vulnerability axioms and their relationships:
\begin{itemize}
    \item When a vulnerable design is exploited by a specific type of vulnerability, it can result in various attack impacts. Consequently, an instance of the `ExploitTarget' class opens the possibility for potential future attacks. To model this relationship, we utilize the following predicate logic:
    \begin{align}
& \exists  x,y: Vulnerability(x)\wedge ExploitTarget(x) \wedge Exploits(x,y)    
\end{align}
        
    \item Every `ExploitTarget' exerts a certain impact on the system, and we encompass all potential impacts caused by a specific vulnerability within the `AttackImpact' class. The object property `hasAttackImpact' establishes the link between an instance of `ExploitTarget' and an instance of `AttackImpact.' This concept is expressed through the following logic:
    \begin{align}
& \exists x,y: ExploitTarget(x) \wedge AttackImpact(x) \wedge hasAttackImpact(x,y)
\end{align}

    \item Every instance of a vulnerability is associated with a CWE-ID that defines the type of weakness, whether it's related to software or hardware. The predicate logic below utilizes the `TargetsCWE' property to establish the connection between a vulnerability instance and its CWE-ID:
    \begin{align}
& \exists x,y: Vulnerability(x) \wedge CWE(x) \wedge TargetsCWE(x,y)    
\end{align}

\end{itemize}

Using the conceptual graph approach, these ontological concepts and interactions are graphically represented in Figure \ref{fig:relationship}, offering a graphical representation for improved comprehension.

\subsection{Automation of the Ontology} 

\begin{figure}[t!]
\centering
\includegraphics[width=\linewidth]{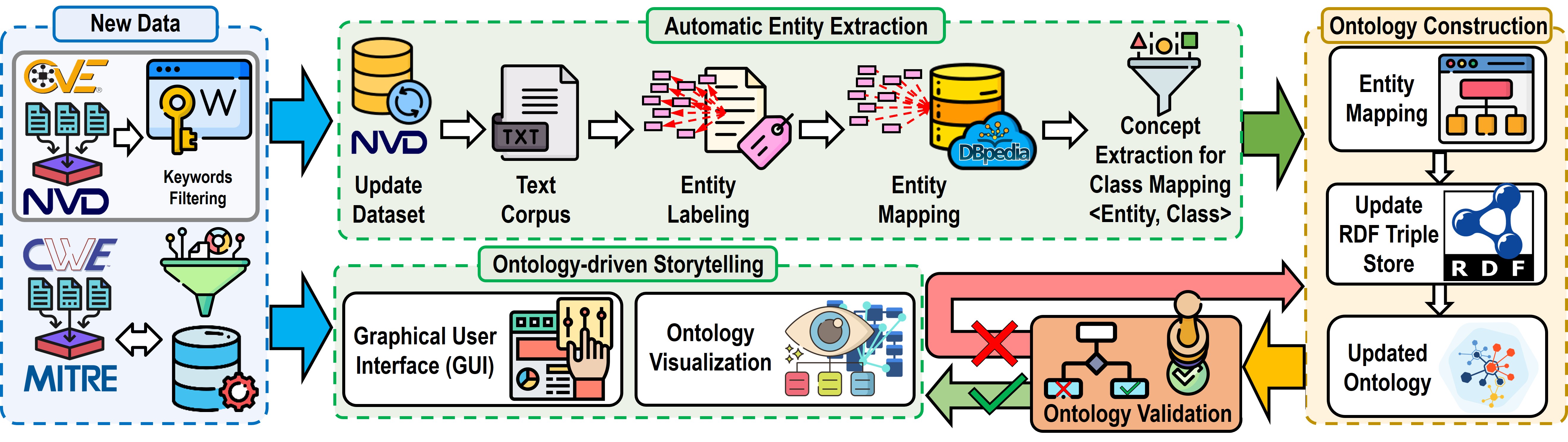}
\caption{Automated Ontology-driven Storytelling Framework.}
\label{fig:OSF}
\end{figure}

We constructed our foundation ontology by using the NVD database from 2010 to 2023. Figure \ref{fig:OSF} illustrates our proposed automated ontology-driven storytelling framework. The first block (from the left to the right) depicts the building block for the acquisition of new data; next, the data flows to the `Automatic Entity Extraction' block to perform the NLP-based data pre-processing required for extraction of entities automatically; following the green arrow, the extracted entities are used in the next step that is the development of an ontology that tells a story about vulnerability patterns. 
Our base ontology is constructed based on the data from the NVD database spanning from 2010 to 2023. Our Ontology-Driven Storytelling Framework (OSF) is described through three main blocks: firstly, `New Data Accumulation,' secondly, `Automatic Entity Extraction,' and finally, `Ontology Construction'—which ultimately narrates the story of vulnerability trends. We illustrate our OSF framework, encompassing the journey from new data accumulation to the visualization of an updated ontology, in Figure \ref{fig:viz}. When new vulnerability data becomes available, our OSF appends this information to the existing NVD database. 

Following a data cleaning process, we proceed with `Automatic Entity Labeling' on the appended data. Subsequently, we establish mappings between the entity labels and DBpedia to enhance the reasoning associated with each entity label. After establishing mappings with the DBPedia corpus, our next step involves `Concept Extraction' for class mapping. This entails assigning each Entity to an appropriate Class. We initiate our ontology construction process with Entity Mapping, followed by updating the RDF Triple Store and subsequently updating the ontology. We map each of the discovered Entities to one of the four aforementioned classes (i.e., `Vulnerability', `CWE', `AttackImpact', and `ExploitTarget'). Next, we proceed to the ontology construction once the concepts are extracted for each class Entity. We initiate our ontology construction process with Entity Mapping, followed by updating the RDF Triple Store and subsequently updating the ontology. We then perform Ontology validation using a python package, Owlready2.0.\footnote{Owlready2 is a python package for creating, modifying, and thus manipulating OWL 2.0 ontologies.} If the updated ontology successfully passes the validation check, we proceed to integrate it into our GUI Framework for ontology-driven storytelling. However, if the updated ontology does not pass the validation check, we backtrack to the RDF Triple Store block and make necessary adjustments until the updated ontology successfully passes the validation check. Moreover, we map the Entity label with DBPedia in order to enhance the logic behind each Entity labeler and create a more consistent knowledge base. We run the `Concept Extraction' process for class mapping after mapping the Entity labels with the DBPedia corpus. This means that we assign each Entity to a Class to complete the mapping process.

A \textit{Vulnerability}, a common weakness enumeration (\textit{CWE}), an \textit{AttackImpact}, and an \textit{ExploitTarget} are the four notions that we have, as was discussed in the preceding section. Every single one of the newly created Entities is assigned to one of these classes. Once we have extracted concepts for each Entity, we will proceed to the next step, the building of the ontology. When we begin the process of building our ontology, we begin by doing entity mapping. We then proceed to update the ontology as well as the RDF Triple Store information. Ontology validation is then carried out with the assistance of a Python tool known as Owlready2.0. If the modified ontology is successful in passing the validation check, we will proceed to incorporate the new ontology into our graphical user interface framework for data-driven storytelling. On the other hand, if the modified ontology is unable to pass the validation check, we will revert to the RDF Triple Store block and make modifications there until the updated ontology can pass the validation evaluation. We will now offer an example to comprehend the newly implemented data extraction procedure better. According to the information presented in Figure \ref{fig:HW-V2W-Map_GUI}, our framework can determine whether or not the NVD database has any new data. The entity concepts are stored as keys in our topic modeling framework, and the information about those concepts is stored as values. This information is then stored in a Python dictionary. After that, we use the Owlready2.0 Python module to map each of the concepts and the relationship that corresponds to it with other concepts. The following is an example of a single entry for NVD:
\begin{verbatim}
nlp_dict=[{
"Vulnerability":"CVE−2020−2020", 
"ExploitTarget":"GoogleChromeOS", 
"AttackImpact":"SpoofingAttack", 
"CWE":"CWE−276"
}]
\end{verbatim}
Following the extraction of concepts from a dictionary, we proceed to develop a vocabulary to enrich our ontology by employing the \textit{<Subject, Predicate, Object>} format, as described below:
\begin{verbatim}
<Vulnerability, hasAttackImpact, AttackImpact>
<ExploitTarget, hasVulnerability, Vulnerability>
<Vulnerability, TargetsCWE, CWE>
\end{verbatim}
This procedure is repeated until all of the new data has been assigned associations, concepts have been captured, and our ontology has been thoroughly updated.

\subsection{Storytelling and Vulnerability Trend Analysis} 

\begin{figure}[t!]
\centering
\includegraphics[width=\linewidth]{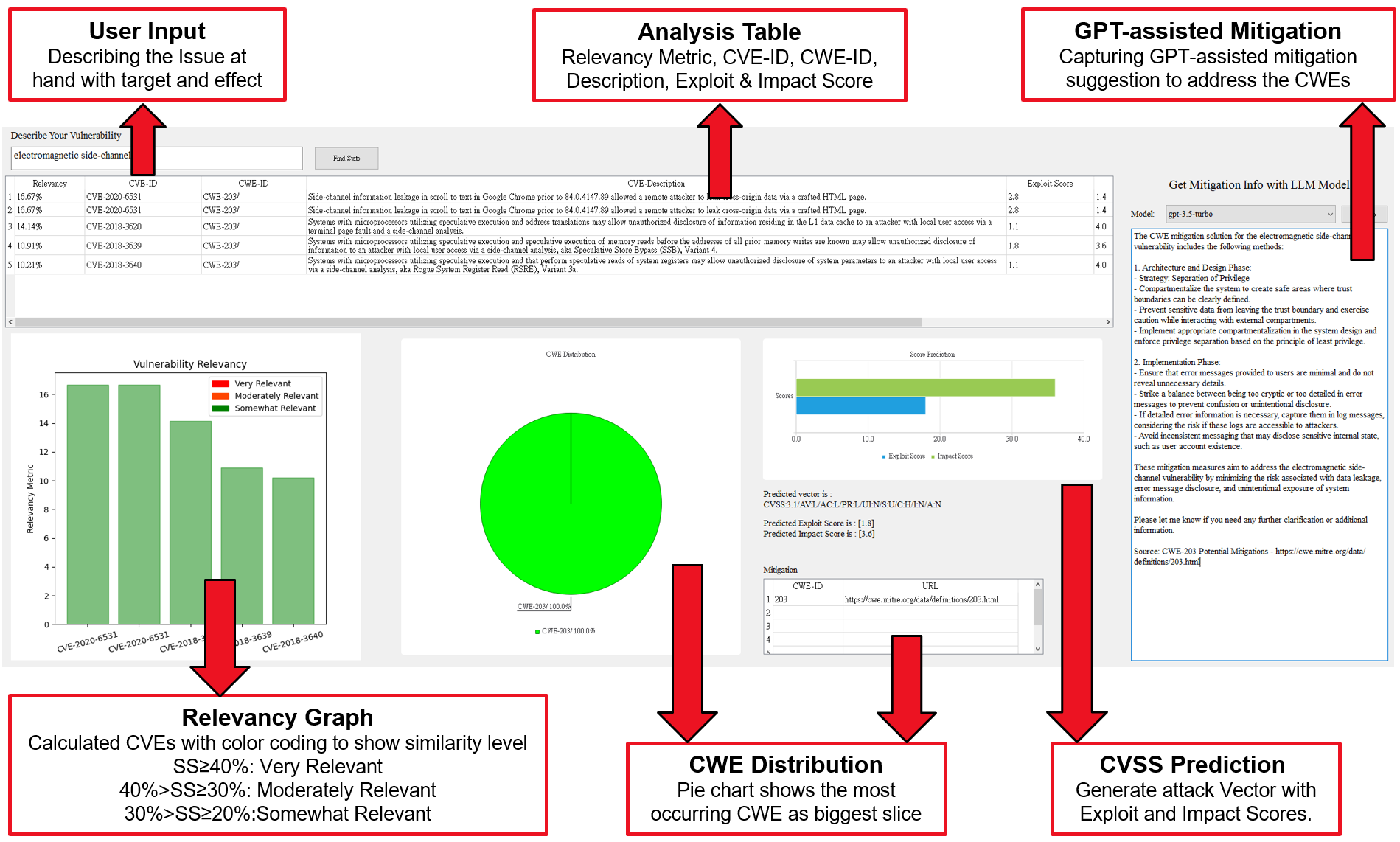}
\caption{HW-V2W-Map GUI.}
\label{fig:HW-V2W-Map_GUI}
\end{figure}

In this section, we discuss the approach taken to record the evaluation process of NVD vulnerabilities. In this study, we establish connections between vulnerabilities and the impact each vulnerability has on victim systems when exploited by an attacker.  We achieve these connections through the utilization of Machine Learning techniques, including linguistic annotations. These techniques involve mapping entities using a previously-made NLP corpus to dynamically relate and attach meaningful contexts to unstructured text corpora. Upon forming this relationship, we proceed to map each vulnerability with its corresponding CVE-ID. Our ontology effectively models the vulnerability information, including all the associated CVE-IDs. To effectively utilize these analysis methods and enhance the visibility of cybersecurity threats, we employed PyQt5 to develop a GUI that offers an interactive user experience, as depicted in Figure \ref{fig:HW-V2W-Map_GUI}. Through our designed GUI, users can input a vulnerability description and receive the data in a well-structured and organized format. We establish links between each CVE and a corresponding CWE and calculate a similarity metric between the input description and each CVE description using cosine similarity. Using our proposed GUI, the user is able to query the knowledge base by providing a description of a vulnerability and obtain the data in a view that is perfectly formatted and organized. In order to determine which subjects are most pertinent to the user input, the framework first establishes a connection between each CVE and a CWE. Next, it calculates the cosine similarity metric between the input description and each CVE description by utilizing the cosine similarity between each vector. Ultimately, an understanding of the type of vulnerability and the distribution of CWEs that could explain the scenario is provided to the user as a result of this function. The proposed procedure for NLP and the computation of cosine similarity is illustrated in Algorithm \ref{alg:topic-modeling}. This process helps identify the most relevant topics pertaining to the input. This functionality enables the user to comprehend the nature of the vulnerability and gain insights into the distribution of CWEs that can help explain the given scenario. 
The next step is to outline the process flow of our proposed framework and graphical user interface.

The GUI undergoes a series of data cleaning steps to process user input explaining vulnerabilities. It leverages the NLP and NLTK framework to eliminate stop words, punctuation, and nonsensical data. Additionally, it tokenizes the input and performs stemming to standardize the description's meaning. To identify similarities in concepts, our proposed Ontology-Driven Storytelling Framework (OSF) calculates cosine similarity to extract relevant occurrences and map the description to existing CVEs. The results from cosine similarity are then presented in the GUI, showcasing the classification of retrieved vulnerabilities and the primary CWE type depicted in the description. Our proposed framework further utilizes the developed ontology to establish connections among these occurrences and present a narrative of their development to the user. This narrative is displayed using OntoSpy to visualize RDF models and facilitate interaction with documentation. In summary, our proposed framework connects these occurrences and illustrates relationships between them by leveraging the established ontology to narrate a developmental story to the user. This storytelling process is enhanced by utilizing OntoSpy to visualize RDF models and provide interactive access to documentation.

\subsection{Vector Prediction} 
In the first step of our vector prediction process, we collected all of the CVSS 3.1 vectors from the list of CVEs that are pertinent to the security of IoT and hardware. This is done in order to forecast the threat level of the data that we are providing. Using the similarity metric from the five vulnerabilities that we match from the CVE data, we determine which vector is the most likely vector from this list of CVEs that are comparable. This is accomplished by determining the feature that occurs the most frequently for each value in the vector. The CVSS vector consists of the following:

\begin{itemize}
\item The attack vector 
\item The complexity of the attack 
\item The privileges that are required 
\item The user interaction 
\item The scope 
\item The impact on confidentiality 
\item The impact on integrity 
\item The impact on availability
\end{itemize}

\begin{figure} [!t]
  \centering
    \includegraphics[width=\textwidth]{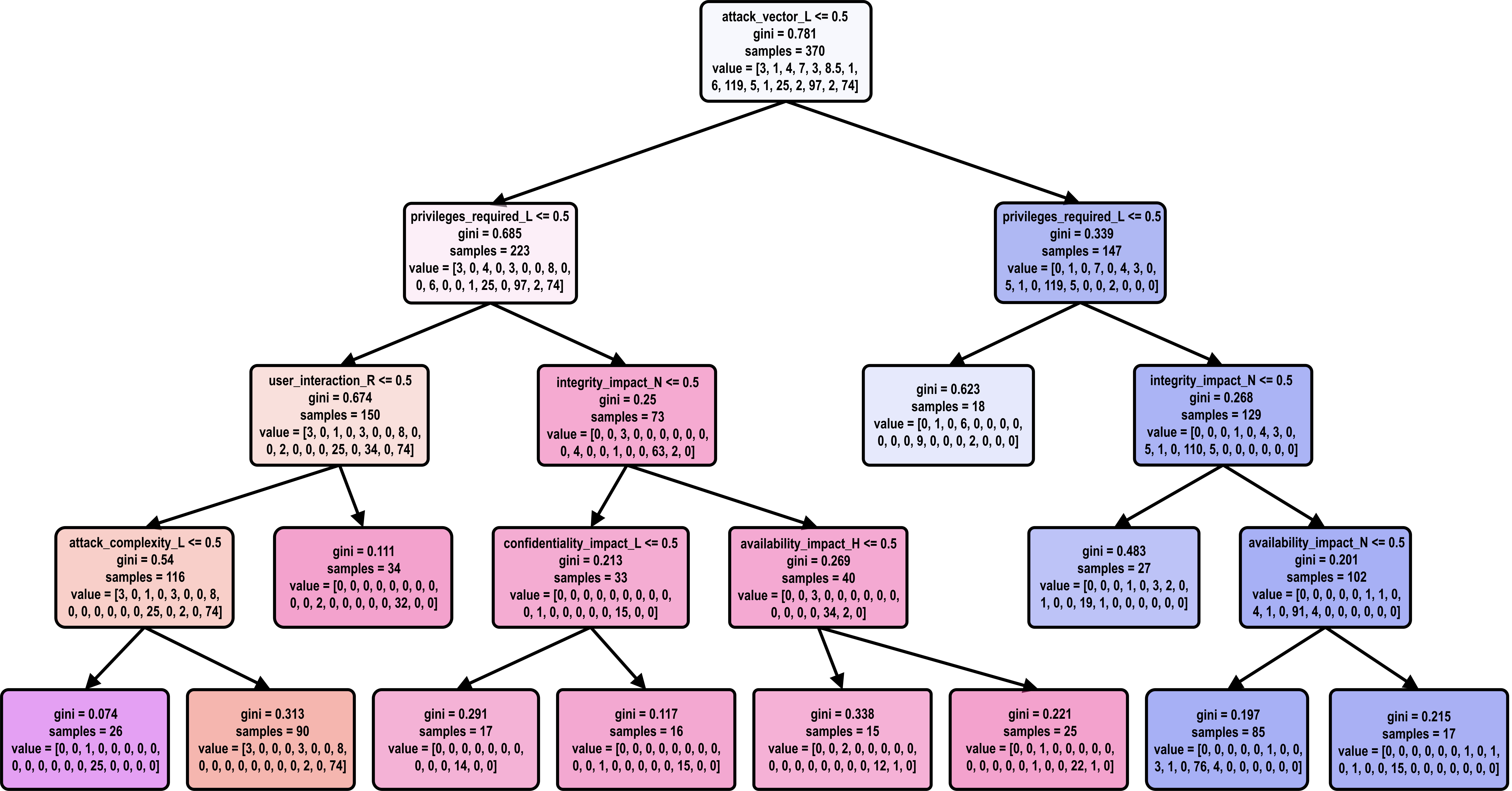}
  \caption{Decision Tree Visualization.}
  \label{fig:decision-tree}
\end{figure}

The approach that we have provided enables us to estimate the impact and exploit score of an input threat, which is made possible by the similarity in the vectors. To train our classifier, we utilize the comprehensive list of vulnerabilities observed in IoT and hardware by employing an ML model called Decision Tree \cite{wu2008top}. After doing some trial and error with several ML models on our experimental data, we eventually decided to select the Decision Tree model as the final model due to its outstanding results compared to other models. 
Utilizing the distributions of the CVE database and the GINI index, our ML approach enables us to identify the most significant or significant attributes. We apply pruning and maximum leaf depth, as well as one-hot-encode for each variation in a feature, to keep our model as accurate and precise as possible. To better illustrate the proposed approach and the assumptions made based on the GINI score, we trimmed our decision tree and presented it in Figure \ref{fig:decision-tree}.

\section{DATA ANALYSIS AND DISCUSSION}

\begin{figure}[!t]
    \centering
    \includegraphics[width=\textwidth]{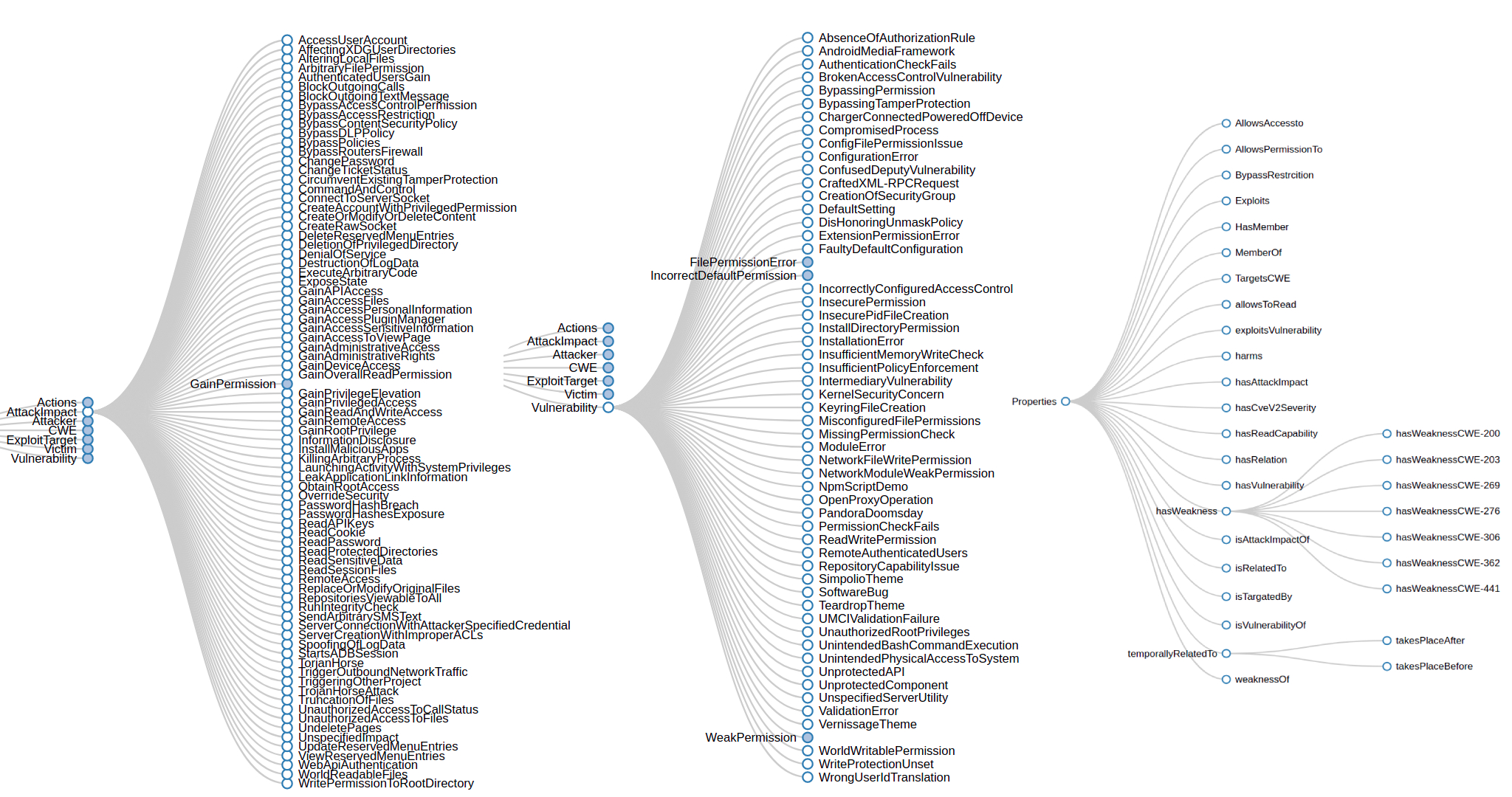}
    \label{fig:onto_visual1}
  \caption{OntoSpy Visuals of the Ontology Interaction.}
  \label{fig:viz}
\end{figure}

The Ontology Web Language (OWL) is utilized to develop our suggested ontology within the Protege 5.5.0 program. A total of 1,460 axioms (logical assertions used to represent the target) including 652 logical and 801 declaration axioms are included in our present ontology. The total number of individuals in the ontology is 518, and there are 252 classes and 32 object properties. We utilized the Ontospy visualization tool in Python in order to accomplish the visualization. Figure \ref{fig:viz} illustrates our suggested ontology, which, due to the limited amount of space available, only expands the classes that are considered to be the most significant.

\begin{figure}[!t]
  \centering
  \begin{subfigure} {\linewidth}
  \centering
    \includegraphics[width=0.6\linewidth]{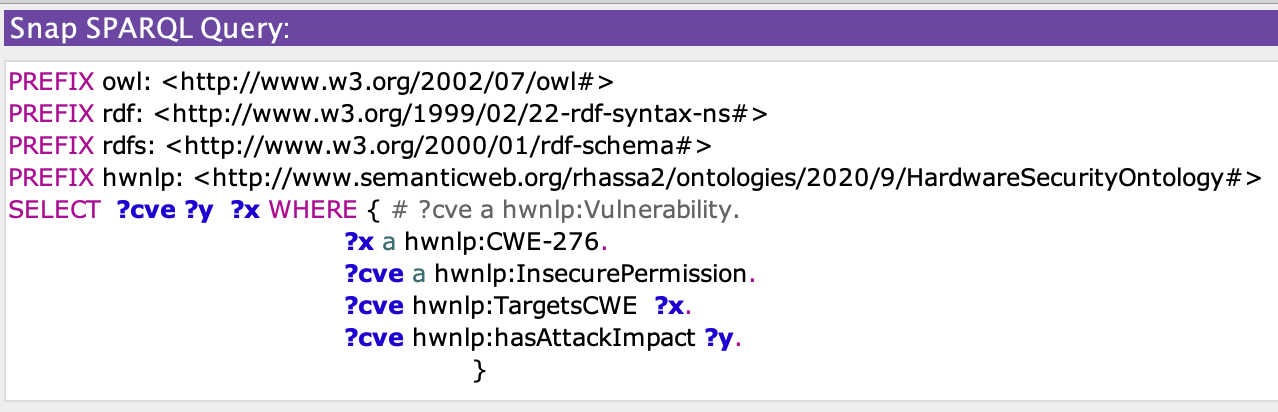}
    \caption{}
    \label{fig:sparql_query}
  \end{subfigure}
  \begin{subfigure} {\linewidth}
  \centering
    \includegraphics[width=0.6\linewidth]{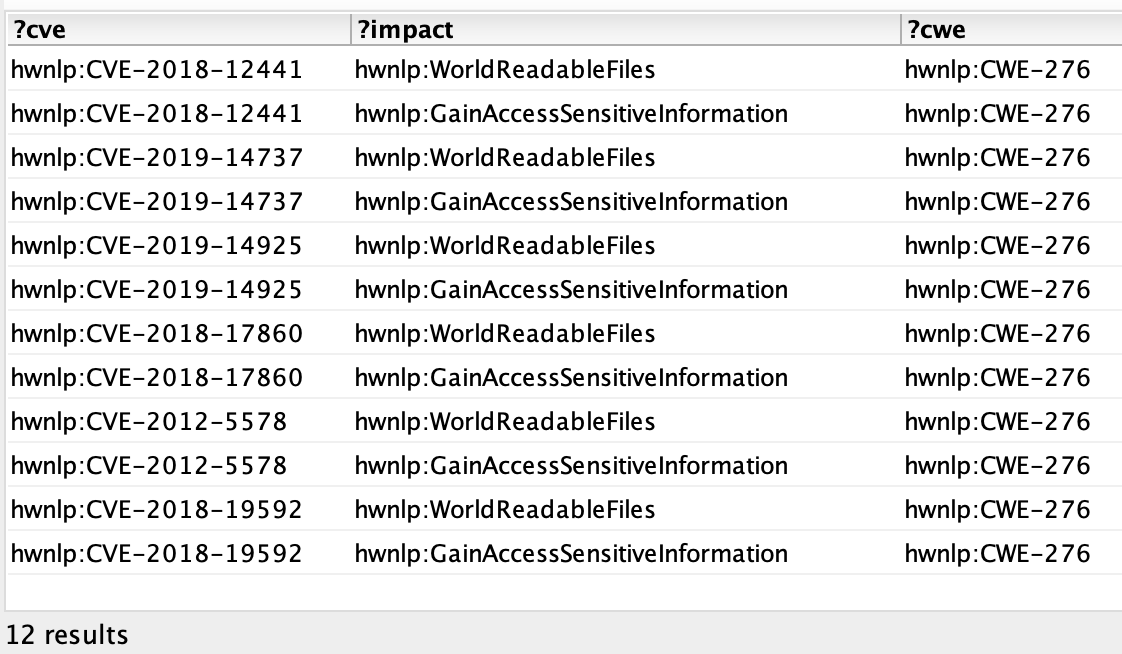}
    \caption{}
    \label{fig:sparql_result}
  \end{subfigure}
  \caption{Attack Impacts of Vulnerabilities related to CWE-276.}
  \label{fig:sparql}
\end{figure}

For our case study, we think about the question, ``What kinds of possible Attack Impacts can a victim system experience for a certain vulnerability that links to a specific CWE-ID (hardware weakness)?'' Before we can answer this question, we need to know the class of the risk instance that was asked about. Second, we need to know what might happen if that weakness is used in a bad way. Each vulnerability is linked to one or more instances of the ``AttackImpact'' class through the ``hasAttackImpact'' connection. This object property will show us all of the possible Attack Impacts for that vulnerability. Finally, each vulnerability is linked to a specific CWE-ID by the ``targetsCWE'' object value. This links our target vulnerability and a specific CWE-ID. The first step in analyzing the attack pattern is asking about a text input corresponding to a certain weakness. When we give our topic modeling framework a sample of text, it uses cosine similarity matching to connect that text to a risk called ``InsecurePermission.'' Our goal is to find Vulnerability, along with its CVE-ID, CWE-ID, and AttackImpact. Figure \ref{fig:sparql_query} displays the SPARQL query used to find the link between the ``InsecurePermission'' vulnerability and the related CWE-276. Figure \ref{fig:sparql_result} shows the 12 answers that the query gives back. Each CVE-ID in this case points to the CWE-276 weakness type.

\subsection{Topic Modeling}
To develop a model that is capable of discovering intricate connections between entities and instances, we use the information obtained from CVE and CWE, as well as the impact that the vulnerability has on the target. We begin by parsing the data and then joining it with the CWE-ID corresponding to each CVE data point. We make use of Algorithm \ref{alg:topic-modeling} and the NLP technique that we have conceptualized. As an illustration, let us take into consideration a CVE such as CVE-2013-4763, which has a description that reads as follows: ``Samsung Galaxy S3/S4 exposes an unprotected component that allows arbitrary SMS text messages without requesting permission.'' Upon completion of our program, we are provided with two outputs, the primary ones being the impacts and the target. By utilizing the SpaCy Linguistic Feature library and correctly matching ``Samsung Galaxy'' as the target, we are able to generate our impact as ``allowing arbitrary SMS text'' and our target as connecting a real-world entity. These precise targets and exploits can be incorporated into our ontology for labeling, as well as into our framework to develop the model.

\subsection{Trends in Vulnerability and Their Relationship with Proposed Technique} \label{sub:Cal_Relation}
Visualizing the data, which gives a summary and connects various pieces of information, assists in constructing a narrative that can serve as a resource for hardware designers to obtain insights into the evolution of vulnerabilities and the relationship between various vulnerabilities. The GUI includes a field for the user to enter information to offer a detailed description of a vulnerability that will be evaluated. Following the completion of the computations by the algorithm, the user is presented with an analytical dashboard that provides detailed information.
The most pertinent CVEs and the specifications associated with them are displayed in a tabular format. A similarity metric is presented in this table, which uses cosine similarity to indicate the degree to which each vulnerability description is related to the user input. Additionally, a CWE, a CVE ID, and a description are displayed, along with the CVSS V3 score, which allows for the access and classification of different levels of threat in terms of exploit and impact score. With the help of color coding, the graph illustrates the relevance of these estimated CVEs. The similarity level is displayed in the GUI by utilizing the following thresholds: A score of 40\% percent or higher indicates that the content is highly relevant. A similarity score between 30\% to 40\% indicates that the content is moderately relevant, while a similarity score below 30\% indicates that the content is somewhat relevant. The following tab displays a pie chart that illustrates the most frequently occurring CWE that has the potential to directly connect to the root cause of the user-described vulnerability. As part of the mitigation procedure, the largest CWE slice will be utilized in order to connect the most similar instances. 

For the sake of describing how we can create the data, we will use an example instance that goes through the complete flow. We are using ``electromagnetic side-channel'' as the user input to the GUI for an observable vulnerability that we are currently dealing with in order to evaluate our proposed ways for obtaining mitigation information. After running the framework using our corpus data, we clean the input string with the function from Algorithm \ref{alg:topic-modeling}. We then calculate the relevance among the vulnerabilities by using a similarity function such as cosine similarity. Finally, we rank the closest match that was found in our database, and the GUI displays the results.
This tabular view presents important results that can be used to identify links between the relevant CVEs. The CVEs that are the most relevant to the user description are listed in this table. In this particular instance, we find that the results indicate a match of 55.69\% with the user-described vulnerability. Additionally, the CVE-IDs and CWE-IDs of the entity that matched our input are displayed in this table. Additionally, the description of the CVE that we matched against the NVD database is listed. In particular, we are dealing with a Java applet threat vulnerability. Our findings from the dataset confirm that this vulnerability is similar to our input. Following this, the proposed framework provides an ML-assisted prediction for the exploit and impact score potentially associated with the user-defined vulnerability.
 
A bar plot that was built with the Matplotlib visualization package is displayed in the GUI, with each bar in the plot representing the similarity measure that was explained earlier. Based on this graph, we can see that the CVE-2020-6531, which is highlighted in green, has the highest similarity score between the two. An explanation of all of the CWEs that are connected to the CVEs is provided in the CWE distribution chart that is included in the pie chart. Additionally, the larger slice is highlighted in the chart. In this particular illustration, we can see one CWE-IDs of this kind, with CWE-203 being the larger slice because it represents the vulnerability known as ``Observable Discrepancy'' The connection between the description and the instance of CVE provides information about how a trend occurs among these similar vulnerabilities. On the other hand, the description of the CWE provides an explanation of what might be the root cause of our input description. These connections will also be added to the final input for our ontology flow to determine the relationship between such entities. 
The proposed framework explains the ML-based severity score vector prediction approach that we discussed earlier. It is composed of three sections, one of which is a horizontal bar graph that illustrates the comparison between exploit and impact scores. We process these vector values through our model to obtain the exploit and impact scores. The predicted severity vector is provided in the text in the following paragraph. In order to evaluate the effectiveness of our ML strategy, we obtained several results, some of which showed an accuracy of up to 98.29\% and a recall score of 90.90\%. All true positive values are returned by this model's precision, yet the accuracy of the values it produces is only 88\%.
Following the presentation of the results, we send the data to the ontology processing. At this stage, we are interested in determining the potential consequences that may arise if an ExploitTarget object is compromised. The ExploitTarget object ``GoogleChromeOs'' is obtained from the topic modeling framework to fulfill user input requirements. We use SPARQL query to accomplish this operation, which allows us to query our ontology to obtain information about the mitigations that are related to the potential attack impact and root cause associated with ``GoogleChromeOs.'' We found that Google Chrome OS is being attacked via a variety of vulnerabilities, and according to our observations, each vulnerability has a corresponding AttackImpact on a machine that is being targeted.

\section{GPT-assisted Mitigation Suggestion}


\begin{figure}[t!]
\centering
\includegraphics[width=\linewidth]{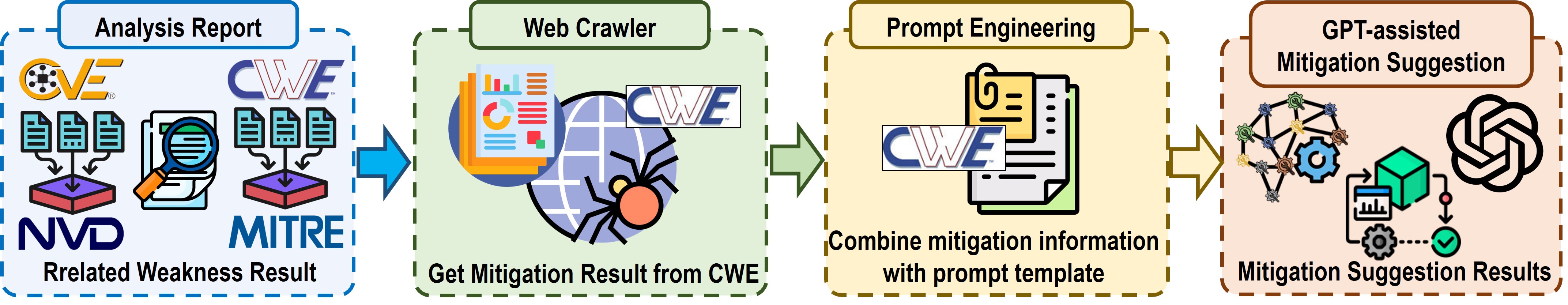}
\caption{GPT-assisted Mitigation Suggestion.}
\label{fig:LLM_Flow_Chart}
\end{figure}

\begin{algorithm}[t!]
\caption{Mitigation Suggestion utilizing OpenAI's GPT LLM}
\label{alg:Get_LLM_Respond}
\footnotesize
\begin{algorithmic}[1]
\State \textbf{Input}: $CWE\_IDs$, $Vulnerability\_Description$, $ModelType$, $\textbf{PromptTemplate}$, $\textbf{WebCrawler}$, $\textbf{Get\_LLM\_Respond}$
\State \textbf{Output}: $Mitigation\_Information$
\State $API\_Key \gets$ Environment Variable \texttt{\$API\_KEY\$}
\State $Mitigation\_Info \gets$ Empty String
\For{$CWE\_ID$ \textbf{in} $CWE\_IDs$}
    \State $Mitigation\_Info \gets Mitigation\_Info + \textbf{WebCrawler}(CWE\_ID)$
\EndFor
\State $\textit{Prompt} \gets \textbf{Prompt\_Template.format}(\textit{Vulnerability\_Description}, \textit{Mitigation\_Info})$
\State \(Mitigation\_Info \gets \textbf{Get\_LLM\_Respond}(\textit{API\_Key}, \textit{ModelType}, \textit{Prompt})\)
\State \textbf{return} \(Mitigation\_Info\)
\end{algorithmic}
\end{algorithm}

After obtaining a list of associated CWE vulnerabilities, the framework provides links and displays them on the GUI, allowing users to access mitigation information easily on the CWE database website. Additionally, the framework introduces an innovative application of Generative Pre-Trained Transformer (GPT) Large Language Models (LLMs) for mitigation suggestions. The flow chart for obtaining GPT-assisted feedback is shown in Figure \ref{fig:LLM_Flow_Chart}. We devised a prompt template for this application, designed to guide the GPT large language model (LLM) in adhering to response rules through a method known as prompt engineering. This approach aims to enable the LLM to generate more specific feedback \cite{white2023prompt, huang2023git}. When utilizing the mitigation suggestion feature, it integrates a web crawler ($\textbf{WebCrawler}$) to fetch the HTML code from specific CWE websites. This web crawler is implemented with BeautifulSoup \cite{richardson2007beautiful}, and this framework uses BeautifulSoup to parse the content of the ``Potential Mitigation'' field. Subsequently, incorporating the user-described vulnerability into the prompt template ($\textbf{PromptTemplate}$).
This framework designed a $\textbf{Get\_LLM\_Respond}$ function through the OpenAI API, and obtains feedback from OpenAI's GPT LLM by passing in the API Key ($API\_Key$), the specified model ($ModelType$), and the content (\textit{Prompt}). Upon receiving the results, the result will be presented on the GUI for user review. The pseudo-code of obtaining LLM feedback from OpenAI's GPT model is shown in Algorithm \ref{alg:Get_LLM_Respond}. 
To show the output of the GPT-assisted Mitigation Suggestion, the ``electromagnetic side-channel'' vulnerability description input mentioned in Section \ref{sub:Cal_Relation} is used. The following information is the output from LLM:
\begin{small}
\begin{Verbatim}[breaklines=true]
The CWE mitigation solution for the electromagnetic side-channel vulnerability includes the following methods:

1. Architecture and Design Phase:
- Strategy: Separation of Privilege
- Compartmentalize the system to create safe areas where trust boundaries can be clearly defined.
- Prevent sensitive data from leaving the trust boundary and exercise caution while interacting with external compartments.
- Implement appropriate compartmentalization in the system design and enforce privilege separation based on the principle of least privilege.

2. Implementation Phase:
- Ensure that error messages provided to users are minimal and do not reveal unnecessary details.
- Strike a balance between being too cryptic or too detailed in error messages to prevent confusion or unintentional disclosure.
- If detailed error information is necessary, capture them in log messages, considering the risk if these logs are accessible to attackers.
- Avoid inconsistent messaging that may disclose sensitive internal state, such as user account existence.

These mitigation measures aim to address the electromagnetic side-channel vulnerability by minimizing the risk associated with data leakage, error message disclosure, and unintentional exposure of system information.

Please let me know if you need any further clarification or additional information.

Source: CWE-203 Potential Mitigations - https://cwe.mitre.org/data/definitions/203.html
\end{Verbatim}
\end{small}

On the other hand, we tested four other vulnerability descriptions, including ``Timing Side Channel'' (Appendix \ref{A:Time-Side}), ``write protection lock bit was left unset''(Appendix \ref{A:write}), ``Administrative access and bypass authentication''(Appendix \ref{A:Administrative}), ``fault inject during the execution''(Appendix \ref{A:fault}). For detailed LLM output, LLM input, and calculation screenshots please check them in the appendix.
This result was obtained using "gpt-3.5-turbo" on December 7, 2023. Please note that the date is specifically mentioned because OpenAI will continue to upgrade or modify the model. From the LLM output results of the ``electromagnetic side-channel'' vulnerability description input, we can observe that through prompt engineering, this reply successfully parsed the HTML code from the CWE webpage and organized the mitigation in an organized manner. In addition, the correct source URL was given as a reference. The detailed LLM input message for this output is in Appendix \ref{A:electro_llm_Input}. This case shows the application potential of GPT LLMs in hardware security. The following is the future work on mitigating suggestion feature of the proposed HW-V2W-Map framework:

\begin{enumerate}
    \item Improve prompt engineering and apply different tasks by selecting different prompt templates.
    \item Provide interactive functions so that users can try and obtain further information from large language models.
    \item Recently, the LLM model has demonstrated the ability to combine image recognition with text information.\cite{waisberg2023gpt}. Providing analysis charts and text information to LLM in the field of hardware network security can potentially improve LLM feedback results.
\end{enumerate}

\section{CONCLUSION}
Our study aimed to enhance the analysis of the risks and consequences associated with NVD by creating an innovative framework for mitigating vulnerabilities, predicting scores, and generating knowledge using an ontology-driven approach. Only plain text input is required for our suggested method, which has the capability of highlighting a variety of pertinent information on the fly. Furthermore, our suggested ontology can quickly identify potential Attack Impacts and Scope, as well as related vulnerabilities for each vulnerability. In addition to this, we have built a GUI that presents all of the information regarding vulnerability and impact analysis in a manner that is both organized and meaningful. To the best of our knowledge, this is the first ontology-driven framework for reducing vulnerabilities, score prediction, and a knowledge-generating system framework that focuses on vulnerabilities in Internet of Things hardware. Moreover, the proposed framework provides mitigation suggestions to address the vulnerabilities utilizing GPT LLMs. In the future, we plan to expand the framework that has been proposed by including other domain concepts.


\bibliographystyle{ACM-Reference-Format}
\bibliography{ref}
\appendix
\section{Obtain HW-V2W-Map}
To obtain HW-V2W-Map, please check this URL: \url{https://gitlab.com/yuzhenglin/HW-V2W-Map}

\subsection{Quick Start}
We provide an executable file for Windows users, which must only be decompressed to execute. The execution steps are as follows:
\begin{enumerate}
    \item Download ``exe\_for\_win\_<yyyymmdd>.zip'' and unzip it.
    \item Set the Open AI API Key to ``OPENAI\_API\_KEY'' in the environment variable.
    \item Double click on ``main\_<yyyymmdd>.exe''
    \item You will see the HW-V2W-Map GUI
\end{enumerate}

\subsection{For Developer}
The following is the current file structure and description of HW-V2W-Map:
\begin{enumerate}
    \item main\_<yyyymmdd>.py: Main function of HW-V2W-Map framework.
    \item prompt\_template.txt: Prompt template for OpenAI's API
    \item requirements.txt: Requirement list of python model.
    \item Modules: Module placement folder.
    \item Database: CVE, CWE, and Topic Model placement folder.
\end{enumerate}

\section{The Detail of GPT-assisted Mitigation Suggestion Experiment}

\subsection{Vulnerability Description: ``electromagnetic side-channel''}
\subsubsection{GPT-assisted Mitigation Suggestion Input}\label{A:electro_llm_Input}
\begin{small}
\begin{Verbatim}[breaklines=true]
- Be concise 
- Be casual unless otherwise specified
- Suggest solutions that I didn't think about — anticipate my needs
- Be accurate and thorough
- Answer immediately. Provide detailed explanations and restate my query in your own words if necessary after giving the answer
- Value good arguments over authorities; the source is irrelevant
- You may use high levels of speculation or prediction; just flag it for me
- No moral lectures
- Discuss safety only when it's crucial and non-obvious
- If your content policy is an issue, provide the closest acceptable response and explain the content policy issue afterward
- Cite sources whenever possible at the end, not inline
- No need to mention your knowledge cutoff
- No need to disclose you're an AI
- If you see the HTML content, extract and organize the information from this HTML code
- List the different phase mitigation method

If I ask for adjustments to the code I provided, only repeat some of my code unnecessarily. Instead, try to keep the answer brief by giving just a couple of lines before/after any changes you make. Multiple code blocks are ok.

Please act as a cybersecurity expert and consultant. I will provide you with a series of questions about cybersecurity.

My question is:
Based on the following vulnerability description and the CWE mitigation information. Feedback on the mitigation solution for me.
Vulnerbility Descreption:electromagnetic side-channel
CWE Mitigation Information(HTML):- CWE 203 - Potential Mitigation:
<div id="Potential_Mitigations"><div class="heading"><span id="script"><a href="javascript:toggleblocksOC('203_Potential_Mitigations');"><img alt="+" border="0" id="ocimg_203_Potential_Mitigations" src="/images/head_more.gif"/></a> </span>Potential Mitigations</div><div class="expandblock" id="oc_203_Potential_Mitigations" name="oc_203_Potential_Mitigations"><div class="detail"><div class="indent"><div id="Grouped"><table border="0" cellpadding="0" cellspacing="0" class="Detail" width="98%">
<tr><td valign="top"><p class="subheading">Phase:  Architecture and Design</p><div class="indent"><p class="suboptheading">Strategy:  Separation of Privilege</p></div><div class="indent"> </div><p><div class="indent">Compartmentalize the system to have "safe" areas where trust 
boundaries can be unambiguously drawn. Do not allow sensitive data to go outside of the trust boundary and always be careful when interfacing with a compartment outside of the safe area. </div></p><div class="indent"> </div><p><div class="indent">Ensure that appropriate compartmentalization is built into the system design, and the compartmentalization allows for and reinforces privilege separation functionality. Architects and designers should rely on the principle of least privilege to decide the appropriate time to use privileges and the time to drop privileges. </div></p><div class="indent"> </div></td></tr>
<tr><td valign="top"><p class="subheading">Phase:  Implementation</p><div class="indent"> </div><p><div class="indent">Ensure that error messages only contain minimal details that are useful to the intended audience and no one else. The messages need to strike the balance between being too cryptic (which can confuse users) or being too detailed (which may reveal more than intended). The messages should not reveal 
the methods that were used to determine the error. Attackers can use detailed information to refine or optimize their original attack, thereby increasing their chances of success. </div></p><div class="indent"> </div><p><div class="indent">If errors must be captured in some detail, record them in log messages, but consider what could occur if the log messages can be viewed by attackers. Highly sensitive information such as passwords should never be saved to log files. </div></p><div class="indent"> </div><p><div class="indent">Avoid inconsistent messaging that might accidentally tip off an attacker about internal state, such as whether a user account exists or not. </div></p><div class="indent"> </div></td></tr>
</table></div></div></div></div></div>

\end{Verbatim}
\end{small}

\subsection{Vulnerability Description:``Timing Side Channel''}\label{A:Time-Side}
\begin{figure} [t!]
  \centering
    \includegraphics[width=\textwidth]{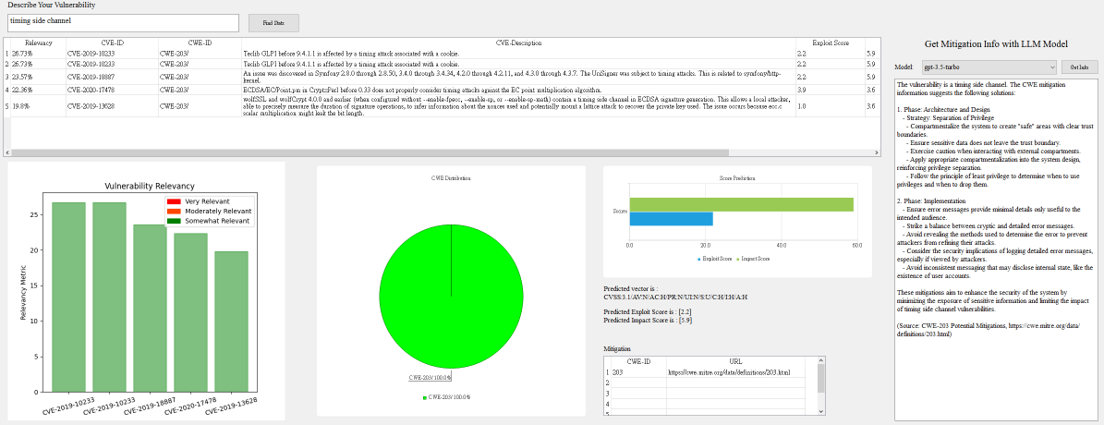}
  \caption{Calculation Result Screenshot of ``Timing Side Channel''}
  \label{fig:time_calc}
\end{figure}

The screenshot of ``Timing Side Channel'' calculation result is shown in Figure \ref{fig:time_calc}.

\subsubsection{GPT-assisted Mitigation Suggestion Output}

\begin{small}
\begin{Verbatim}[breaklines=true]
The vulnerability is a timing side channel. The CWE mitigation information suggests the following solutions:

1. Phase: Architecture and Design
   - Strategy: Separation of Privilege
     - Compartmentalize the system to create "safe" areas with clear trust boundaries.
     - Ensure sensitive data does not leave the trust boundary.
     - Exercise caution when interacting with external compartments.
     - Apply appropriate compartmentalization into the system design, reinforcing privilege separation.
     - Follow the principle of least privilege to determine when to use privileges and when to drop them.

2. Phase: Implementation
   - Ensure error messages provide minimal details only useful to the intended audience.
   - Strike a balance between cryptic and detailed error messages.
   - Avoid revealing the methods used to determine the error to prevent attackers from refining their attacks.
   - Consider the security implications of logging detailed error messages, especially if viewed by attackers.
   - Avoid inconsistent messaging that may disclose internal state, like the existence of user accounts.

These mitigations aim to enhance the security of the system by minimizing the exposure of sensitive information and limiting the impact of timing side channel vulnerabilities.

(Source: CWE-203 Potential Mitigations, https://cwe.mitre.org/data/definitions/203.html)

\end{Verbatim}
\end{small}

\subsubsection{GPT-assisted Mitigation Suggestion Intput}
\begin{small}
\begin{Verbatim}[breaklines=true]
<Prompt Template is Same as ''electromagnetic side-channel'' Input>

My question is:
Based on the following vulnerability description and the CWE mitigation information. Feedback on the mitigation solution for me.
Vulnerbility Descreption:Timing Side Channel
CWE Mitigation Information(HTML):- CWE 203 - Potential Mitigation:
<div id="Potential_Mitigations"><div class="heading"><span id="script"><a href="javascript:toggleblocksOC('203_Potential_Mitigations');"><img alt="+" border="0" id="ocimg_203_Potential_Mitigations" src="/images/head_more.gif"/></a> </span>Potential Mitigations</div><div class="expandblock" id="oc_203_Potential_Mitigations" name="oc_203_Potential_Mitigations"><div class="detail"><div class="indent"><div id="Grouped"><table border="0" cellpadding="0" cellspacing="0" class="Detail" width="98%">
<tr><td valign="top"><p class="subheading">Phase:  Architecture and Design</p><div class="indent"><p class="suboptheading">Strategy:  Separation of Privilege</p></div><div class="indent"> </div><p><div class="indent">Compartmentalize the system to have "safe" areas where trust 
boundaries can be unambiguously drawn. Do not allow sensitive data to go outside of the trust boundary and always be careful when interfacing with a compartment outside of the safe area. </div></p><div class="indent"> </div><p><div class="indent">Ensure that appropriate compartmentalization is built into the system design, and the compartmentalization allows for and reinforces privilege separation functionality. Architects and designers should rely on the principle of least privilege to decide the appropriate time to use privileges and the time to drop privileges. </div></p><div class="indent"> </div></td></tr>
<tr><td valign="top"><p class="subheading">Phase:  Implementation</p><div class="indent"> </div><p><div class="indent">Ensure that error messages only contain minimal details that are useful to the intended audience and no one else. The messages need to strike the balance between being too cryptic (which can confuse users) or being too detailed (which may reveal more than intended). The messages should not reveal 
the methods that were used to determine the error. Attackers can use detailed information to refine or optimize their original attack, thereby increasing their chances of success. </div></p><div class="indent"> </div><p><div class="indent">If errors must be captured in some detail, record them in log messages, but consider what could occur if the log messages can be viewed by attackers. Highly sensitive information such as passwords should never be saved to log files. </div></p><div class="indent"> </div><p><div class="indent">Avoid inconsistent messaging that might accidentally tip off an attacker about internal state, such as whether a user account exists or not. </div></p><div class="indent"> </div></td></tr>
</table></div></div></div></div></div>

\end{Verbatim}
\end{small}

\subsection{Vulnerability Description:``write protection lock bit was left unset''}\label{A:write}
The screenshot of ``write protection lock bit was left unset'' calculation result is shown in Figure \ref{fig:admin_calc}.
\begin{figure} [t!]
  \centering
    \includegraphics[width=\textwidth]{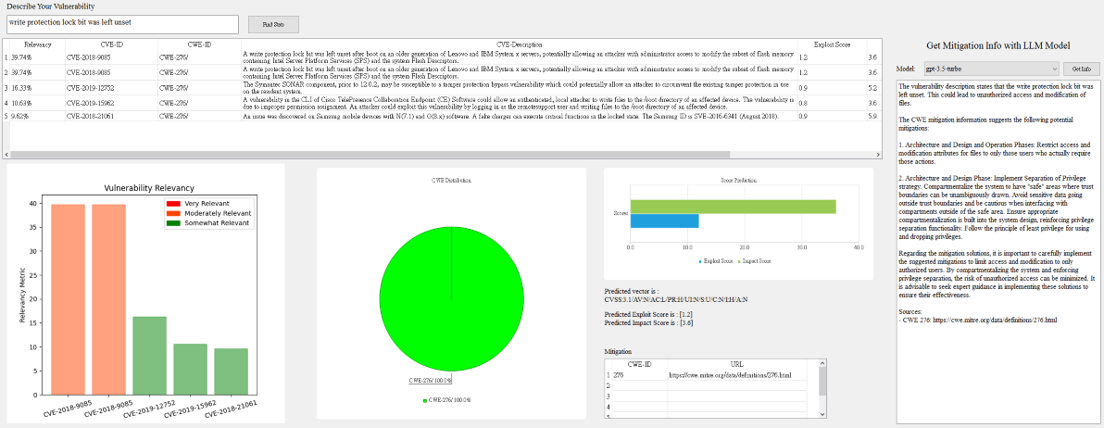}
  \caption{Calculation Result Screenshot of ``write protection lock bit was left unset''}
  \label{fig:write_calc}
\end{figure}

\subsubsection{GPT-assisted Mitigation Suggestion Output}
\begin{small}
\begin{Verbatim}[breaklines=true]
The vulnerability description states that the write protection lock bit was left unset. This could lead to unauthorized access and modification of files. 

The CWE mitigation information suggests the following potential mitigations:

1. Architecture and Design and Operation Phases: Restrict access and modification attributes for files to only those users who actually require those actions.

2. Architecture and Design Phase: Implement Separation of Privilege strategy. Compartmentalize the system to have "safe" areas where trust boundaries can be unambiguously drawn. Avoid sensitive data going outside trust boundaries and be cautious when interfacing with compartments outside of the safe area. Ensure appropriate compartmentalization is built into the system design, reinforcing privilege separation functionality. Follow the principle of least privilege for using and dropping privileges.

Regarding the mitigation solutions, it is important to carefully implement the suggested mitigations to limit access and modification to only authorized users. By compartmentalizing the system and enforcing privilege separation, the risk of unauthorized access can be minimized. It is advisable to seek expert guidance in implementing these solutions to ensure their effectiveness.

Sources: 
- CWE 276: https://cwe.mitre.org/data/definitions/276.html

\end{Verbatim}
\end{small}

\subsubsection{GPT-assisted Mitigation Suggestion Intput}
\begin{small}
\begin{Verbatim}[breaklines=true]
<Prompt Template is Same as "electromagnetic side-channel" Input>

Based on the following vulnerability description and the CWE mitigation information. Feedback on the mitigation solution for me.
Vulnerbility Descreption:write protection lock bit was left unset
CWE Mitigation Information(HTML):- CWE 276 - Potential Mitigation:
<div id="Potential_Mitigations"><div class="heading"><span id="script"><a href="javascript:toggleblocksOC('276_Potential_Mitigations');"><img alt="+" border="0" id="ocimg_276_Potential_Mitigations" src="/images/head_more.gif"/></a> </span>Potential Mitigations</div><div class="expandblock" id="oc_276_Potential_Mitigations" name="oc_276_Potential_Mitigations"><div class="detail"><div class="indent"><div id="Grouped"><table border="0" cellpadding="0" cellspacing="0" class="Detail" width="98%">
<tr><td valign="top"><p class="subheading">Phases:  Architecture and Design; Operation</p><div class="indent">The architecture needs to access and modification attributes for files to only those users who actually require those actions. </div></td></tr>
<tr><td valign="top"><p class="subheading">Phase:  Architecture and Design</p><div class="indent"><p class="suboptheading">Strategy:  Separation of Privilege</p></div><div class="indent"> </div><p><div class="indent">Compartmentalize the system to have "safe" areas where trust 
boundaries can be unambiguously drawn. Do not allow sensitive data to go outside of the trust boundary and always be careful when interfacing with a compartment outside of the safe area. </div></p><div class="indent"> </div><p><div class="indent">Ensure that appropriate compartmentalization is built into the system design, and the compartmentalization allows for and reinforces privilege separation functionality. Architects and designers should rely on the principle of least privilege to decide the appropriate time to use privileges and the time to drop privileges. </div></p><div class="indent"> </div></td></tr>
</table></div></div></div></div></div>
\end{Verbatim}
\end{small}

\subsection{Vulnerability Description:``Administrative access and bypass authentication''}\label{A:Administrative}
\begin{figure} [t!]
  \centering
    \includegraphics[width=\textwidth]{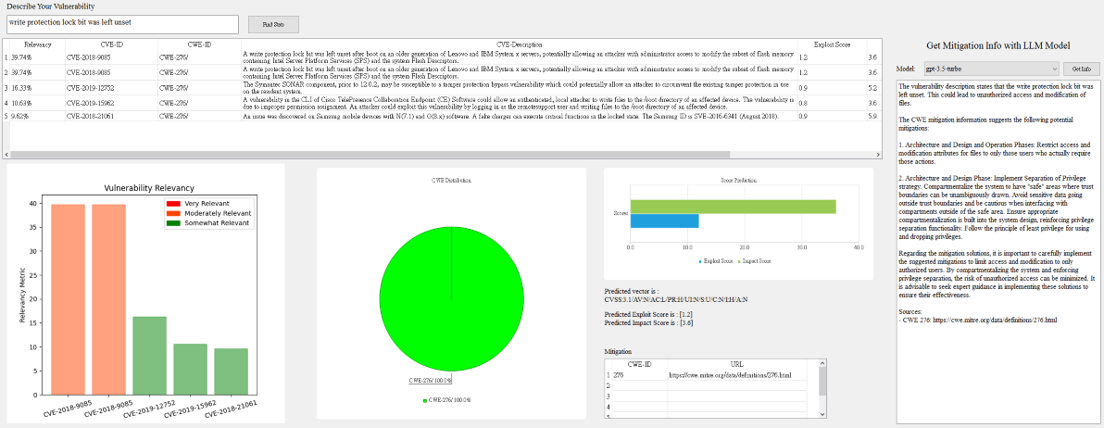}
  \caption{The screenshot of ``Administrative access and bypass authentication'' calculation result}
  \label{fig:admin_calc}
\end{figure}

The screenshot of ``Administrative access and bypass authentication'' calculation result is shown in Figure \ref{fig:admin_calc}.

\subsubsection{GPT-assisted Mitigation Suggestion Output}
\begin{small}
\begin{Verbatim}[breaklines=true]
The potential mitigation solution for the vulnerability of administrative access and bypass authentication includes the following steps:

1. Architecture and Design phase: Restrict access to file attributes only to users who require those actions.
2. Architecture and Design phase: Implement the strategy of Separation of Privilege to compartmentalize the system and establish trust boundaries.
3. Architecture and Design phase: Ensure that sensitive data does not exit the trust boundary and exercise caution when interfacing with external compartments.
4. Architecture and Design phase: Incorporate appropriate compartmentalization into the system design, reinforcing privilege separation functionality.
5. Architecture and Design phase: Follow the principle of least privilege to determine when to use privileges and when to drop them.

These mitigation measures will help prevent unauthorized administrative access and bypassing of authentication.

Source: CWE-276 Potential Mitigations (https://cwe.mitre.org/data/definitions/276.html)

\end{Verbatim}
\end{small}

\subsubsection{GPT-assisted Mitigation Suggestion Intput}
\begin{small}
\begin{Verbatim}[breaklines=true]
<Prompt Template is Same as "electromagnetic side-channel" Input>

My question is:
Based on the following vulnerability description and the CWE mitigation information. Feedback on the mitigation solution for me.
Vulnerbility Descreption:Administrative access and bypass authentication
CWE Mitigation Information(HTML):- CWE 276 - Potential Mitigation:
<div id="Potential_Mitigations"><div class="heading"><span id="script"><a href="javascript:toggleblocksOC('276_Potential_Mitigations');"><img alt="+" border="0" id="ocimg_276_Potential_Mitigations" src="/images/head_more.gif"/></a> </span>Potential Mitigations</div><div class="expandblock" id="oc_276_Potential_Mitigations" name="oc_276_Potential_Mitigations"><div class="detail"><div class="indent"><div id="Grouped"><table border="0" cellpadding="0" cellspacing="0" class="Detail" width="98%">
<tr><td valign="top"><p class="subheading">Phases:  Architecture and Design; Operation</p><div class="indent">The architecture needs to access and modification attributes for files to only those users who actually require those actions. </div></td></tr>
<tr><td valign="top"><p class="subheading">Phase:  Architecture and Design</p><div class="indent"><p class="suboptheading">Strategy:  Separation of Privilege</p></div><div class="indent"> </div><p><div class="indent">Compartmentalize the system to have "safe" areas where trust 
boundaries can be unambiguously drawn. Do not allow sensitive data to go outside of the trust boundary and always be careful when interfacing with a compartment outside of the safe area. </div></p><div class="indent"> </div><p><div class="indent">Ensure that appropriate compartmentalization is built into the system design, and the compartmentalization allows for and reinforces privilege separation functionality. Architects and designers should rely on the principle of least privilege to decide the appropriate time to use privileges and the time to drop privileges. </div></p><div class="indent"> </div></td></tr>
</table></div></div></div></div></div>

\end{Verbatim}
\end{small}

\subsection{Vulnerability Description:``fault inject during the execution''} \label{A:fault}

\begin{figure} [t!]
  \centering
    \includegraphics[width=\textwidth]{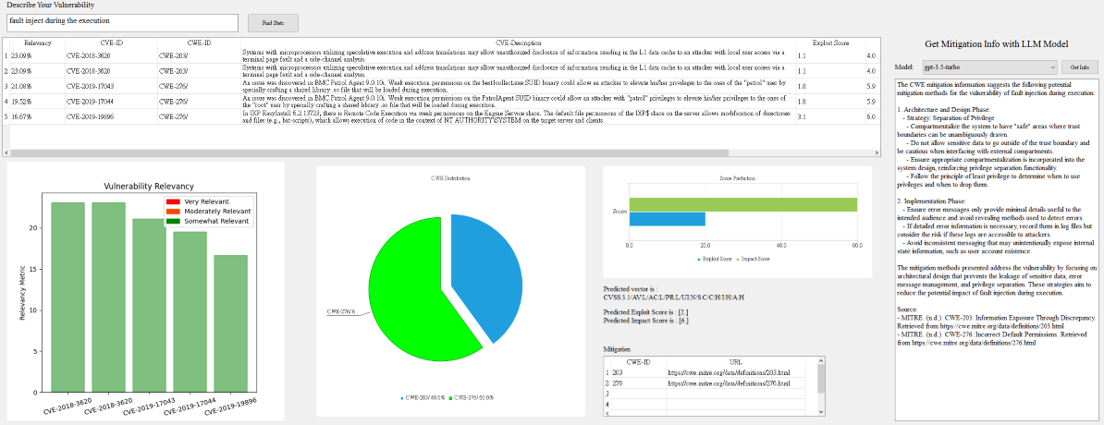}
  \caption{The screenshot of `fault inject during the execution'' calculation result}
  \label{fig:fault_calc}
\end{figure}

The screenshot of ``fault inject during the execution'' calculation result is shown in Figure \ref{fig:fault_calc}.

\subsubsection{GPT-assisted Mitigation Suggestion Output}
\begin{small}
\begin{Verbatim}[breaklines=true]
The CWE mitigation information suggests the following potential mitigation methods for the vulnerability of fault injection during execution:

1. Architecture and Design Phase:
   - Strategy: Separation of Privilege
     - Compartmentalize the system to have "safe" areas where trust boundaries can be unambiguously drawn.
     - Do not allow sensitive data to go outside of the trust boundary and be cautious when interfacing with external compartments.
     - Ensure appropriate compartmentalization is incorporated into the system design, reinforcing privilege separation functionality.
     - Follow the principle of least privilege to determine when to use privileges and when to drop them.

2. Implementation Phase:
   - Ensure error messages only provide minimal details useful to the intended audience and avoid revealing methods used to detect errors.
   - If detailed error information is necessary, record them in log files but consider the risk if these logs are accessible to attackers.
   - Avoid inconsistent messaging that may unintentionally expose internal state information, such as user account existence.

The mitigation methods presented address the vulnerability by focusing on architectural design that prevents the leakage of sensitive data, error message management, and privilege separation. These strategies aim to reduce the potential impact of fault injection during execution.

Source: 
- MITRE. (n.d.). CWE-203: Information Exposure Through Discrepancy. Retrieved from https://cwe.mitre.org/data/definitions/203.html
- MITRE. (n.d.). CWE-276: Incorrect Default Permissions. Retrieved from https://cwe.mitre.org/data/definitions/276.html
\end{Verbatim}
\end{small}

\subsubsection{GPT-assisted Mitigation Suggestion Intput}
\begin{small}
\begin{Verbatim}[breaklines=true]
<Prompt Template is Same as "electromagnetic side-channel" Input>

My question is:
Based on the following vulnerability description and the CWE mitigation information. Feedback on the mitigation solution for me.
Vulnerbility Descreption:fault inject during the execution
CWE Mitigation Information(HTML):- CWE 203 - Potential Mitigation:
<div id="Potential_Mitigations"><div class="heading"><span id="script"><a href="javascript:toggleblocksOC('203_Potential_Mitigations');"><img alt="+" border="0" id="ocimg_203_Potential_Mitigations" src="/images/head_more.gif"/></a> </span>Potential Mitigations</div><div class="expandblock" id="oc_203_Potential_Mitigations" name="oc_203_Potential_Mitigations"><div class="detail"><div class="indent"><div id="Grouped"><table border="0" cellpadding="0" cellspacing="0" class="Detail" width="98%">
<tr><td valign="top"><p class="subheading">Phase:  Architecture and Design</p><div class="indent"><p class="suboptheading">Strategy:  Separation of Privilege</p></div><div class="indent"> </div><p><div class="indent">Compartmentalize the system to have "safe" areas where trust 
boundaries can be unambiguously drawn. Do not allow sensitive data to go outside of the trust boundary and always be careful when interfacing with a compartment outside of the safe area. </div></p><div class="indent"> </div><p><div class="indent">Ensure that appropriate compartmentalization is built into the system design, and the compartmentalization allows for and reinforces privilege separation functionality. Architects and designers should rely on the principle of least privilege to decide the appropriate time to use privileges and the time to drop privileges. </div></p><div class="indent"> </div></td></tr>
<tr><td valign="top"><p class="subheading">Phase:  Implementation</p><div class="indent"> </div><p><div class="indent">Ensure that error messages only contain minimal details that are useful to the intended audience and no one else. The messages need to strike the balance between being too cryptic (which can confuse users) or being too detailed (which may reveal more than intended). The messages should not reveal 
the methods that were used to determine the error. Attackers can use detailed information to refine or optimize their original attack, thereby increasing their chances of success. </div></p><div class="indent"> </div><p><div class="indent">If errors must be captured in some detail, record them in log messages, but consider what could occur if the log messages can be viewed by attackers. Highly sensitive information such as passwords should never be saved to log files. </div></p><div class="indent"> </div><p><div class="indent">Avoid inconsistent messaging that might accidentally tip off an attacker about internal state, such as whether a user account exists or not. </div></p><div class="indent"> </div></td></tr>
</table></div></div></div></div></div>

- CWE 276 - Potential Mitigation:
<div id="Potential_Mitigations"><div class="heading"><span id="script"><a href="javascript:toggleblocksOC('276_Potential_Mitigations');"><img alt="+" border="0" id="ocimg_276_Potential_Mitigations" src="/images/head_more.gif"/></a> </span>Potential Mitigations</div><div class="expandblock" id="oc_276_Potential_Mitigations" name="oc_276_Potential_Mitigations"><div class="detail"><div class="indent"><div id="Grouped"><table border="0" cellpadding="0" cellspacing="0" class="Detail" width="98%">
<tr><td valign="top"><p class="subheading">Phases:  Architecture and Design; Operation</p><div class="indent">The architecture needs to access and modification attributes for files to only those users who actually require those actions. </div></td></tr>
<tr><td valign="top"><p class="subheading">Phase:  Architecture and Design</p><div class="indent"><p class="suboptheading">Strategy:  Separation of Privilege</p></div><div class="indent"> </div><p><div class="indent">Compartmentalize the system to have "safe" areas where trust 
boundaries can be unambiguously drawn. Do not allow sensitive data to go outside of the trust boundary and always be careful when interfacing with a compartment outside of the safe area. </div></p><div class="indent"> </div><p><div class="indent">Ensure that appropriate compartmentalization is built into the system design, and the compartmentalization allows for and reinforces privilege separation functionality. Architects and designers should rely on the principle of least privilege to decide the appropriate time to use privileges and the time to drop privileges. </div></p><div class="indent"> </div></td></tr>
</table></div></div></div></div></div>
\end{Verbatim}
\end{small}

\end{document}